\documentclass[prb,twocolumn,floatfix,showpacs,superscriptaddress]{revtex4-1}
\usepackage{epsfig,subfigure,amssymb,amscd,amsmath,graphicx,amsfonts,mathtools,pbox,amssymb}
\usepackage{times,longtable,math dots,hyperref}
\usepackage{tikz}
\usepackage{rotating}
\usetikzlibrary{decorations.pathreplacing,patterns,calc}

\newcommand{\ba}{\begin{align}}
\newcommand{\ea}{\end{align}}
\newcommand{\braket}[2]{\left\langle #1\middle|#2\right\rangle}
\newcommand{\ketbra}[2]{\left|#1\vphantom{#2}\right\rangle \left\langle \vphantom{#1}#2\right|}
\newcommand{\ket}[1]{     |    \,    #1    \rangle}

\newcommand{\RR}{\mathbb{R}}

\usepackage{tabularx}

\newcommand{\SecRef}[1]{Sec. \ref{#1}}


\newcommand{\PsiCFL}{\psi_{\mathrm{CF}}}
\newcommand{\ie}{\emph{i.e.}~}
\newcommand{\eg}{\emph{e.g.}~}

\renewcommand{\i}{\imath}
\newcommand{\elliptic}[3]{\vartheta_{#1}\!\left(\left.#2\vphantom{#3}\right|#3\right)}

\newcommand{\ellipticgeneralized}[4]{\vartheta\left[\begin{array}{c} #1\\ #2\end{array}\right]\left(\left.#3\vphantom{#4}\right|#4\right)}

\begin{document}

\title{Trial wave functions for a Composite Fermi liquid on a torus}

\author{M. Fremling}
\affiliation{Department of Theoretical Physics, Maynooth University, Ireland.}
\author{N. Moran}
\affiliation{Department of Theoretical Physics, Maynooth University, Ireland.}
\author{J. K. Slingerland}
\affiliation{Department of Theoretical Physics, Maynooth University, Ireland.}
\affiliation{Dublin Institute for Advanced  Studies, School of Theoretical  Physics, 10 Burlington Rd, Dublin, Ireland.}
\affiliation{Rudolf Peierls Centre for Theoretical Physics, 1 Keble Road, Oxford OX1 3NP, United Kingdom.}
\author{S. H. Simon}
\affiliation{Rudolf Peierls Centre for Theoretical Physics, 1 Keble Road, Oxford OX1 3NP, United Kingdom.}

\begin{abstract}
We study the two-dimensional electron gas in a magnetic field at filling fraction $\nu=\frac{1}{2}$. At this filling the system is in a gapless state which can be interpreted as a Fermi liquid of composite fermions. We construct trial wave functions for the system on a torus, based on this idea, and numerically compare these to exact wave functions for small systems found by exact diagonalization. We find that the trial wave functions give an excellent description of the ground state of the system, as well as its charged excitations, in all momentum sectors. We analyze the dispersion of the composite fermions and the Berry phase associated with dragging a single fermion   
around the Fermi surface and comment on the implications of our results for the current debate on whether composite fermions are Dirac fermions.    
\end{abstract}


\date{\today} \maketitle

\section{Introduction}

In the study of the fractional quantum Hall effect, the gapless state at filling fraction $\nu=\frac12$ has enjoyed a special status.
This state can be described in terms of composite fermions\cite{Jain_CF} (CFs) that are moving in a vanishing effective magnetic field, which form a Fermi liquid state\cite{Halperin93,Stern_1999}.
Recently, there has been renewed interest in this state as it has been proposed that the composite fermions could be Dirac fermions\cite{Son2015,Metlitski_2016} and are related to surface states of a 3D topological insulator\cite{Wang_2015_PRX,Wang_2016_PRX,Wang_2016_PRB}.
Recent advances in DMRG for fractional quantum Hall states have shown the
emergence of a Fermi disc\cite{Geraedts_2016}. While DMRG is a powerful general
purpose method, systems based on composite fermions have traditionally been
described using trial wave functions based directly on Slater determinants for
the composite fermions, combined with a flux attachment factor implementing most
of the strong correlations of the actual electrons. These wave functions have
been fundamental to the success of the composite fermion paradigm for fractional
quantum Hall states and allow for a great deal of physical intuition about
non-interacting fermions to be harnessed in the description of the strongly
interacting low energy spectrum.  The aim of this paper is to carefully revisit
and test the composite fermion construction on a torus. This geometry is very
 suited to the study of Fermi liquid type behavior, for example the emergence of Fermi discs, even at relatively small system sizes, but has been difficult to study in the past, because it was difficult to evaluate the natural CF trial wave functions. 

Trial wave functions for the half-filled system based on the composite fermion idea have been studied before. An extensive study  was done by Rezayi and Read in the spherical geometry\cite{Rezayi_1994}. 
The spherical CF construction works very well but is more similar in spirit to atomic physics than to the physics of a Fermi liquid, especially at small system sizes. The organizing quantum number on the sphere is angular momentum, rather than momentum, and for system sizes accessible by exact diagonalization, the lowest energy states at half filling are described by a shell model, with accompanying Hund's rules. 
On a torus, the states of the electron system are organized by momentum and one expects to find ground states which are Fermi discs of composite fermions even at system sizes accessible to exact diagonalization. The torus also has the advantage that different ground states with the same filling fraction all occur at exactly the same magnetic flux and not shifted with respect to each other, as is the case on the sphere. This allows for example the implementation of particle-hole duality as a symmetry of the ground state of exactly half filled systems. Therefore the torus provides a natural setting  for study of the half-filled system.  
Some earlier work on the half-filled system on a torus does exist.  CF trial wave functions like the ones we study here were proposed already in 1994 by Read\cite{Read94,Read96b},
although he did not explicitly specify a toroidal geometry. A notable numerical
study on the torus is Ref.~\onlinecite{Rezayi00}. In that paper, the focus was
on the transition between the CF Fermi liquid and Pfaffian state. Trial wave
functions for composite fermion states were only considered for special momentum
sectors and on the square torus geometry. It appears that there was also interesting unpublished numerical work by Haldane and collaborators, some of which is briefly described in Ref.~\onlinecite{Chari97}.
More recent numerical work in Refs.~\onlinecite{Shao2015} and \onlinecite{Wang2017} use trial wave functions which are not necessarily fully in the LLL.
Other work focuses on approximate LLL-projections for the torus\cite{Pu2017}, generalizing the Jain-Kamilla projection for spherical and planar systems to the torus - this has not been developed for the half filled system so far.

The work presented here complements and extends these works, considering all
momentum sectors, finding the global ground state, studying the CF dispersion, investigating the role of particle-hole symmetry and other discrete symmetries in the half filled state and considering tori of different geometries. We also consider the charge excitations of the CF liquid.
We use energy projection\cite{Fremling_16} to obtain a controlled approximation of the LLL projection for system sizes up to those reached by exact diagonalization.

This paper is organized as follows.
In \SecRef{sec:Wave_Func} we introduce the composite fermion wave functions for a half filled Landau level on the torus and we discuss the relevant symmetries and momentum sectors of the toroidal system.

In \SecRef{sec:Landscapes}, we  determine to what extent a model of non-interacting CFs filling a Fermi disc can reproduce the actual energy landscape seen in exact diagonalization of finite electron systems. We find strong qualitative agreement between the energy landscapes from ED and those produced by non-interacting CFs.

We continue, in \SecRef{sec:PH_inv}, with an investigation of the effects of particle-hole and orbital inversion transformations on the projected CF states.
These transformations do not commute with the center of mass momentum and therefore mix the momentum sectors (in essentially the same way).
While our trial states are not exactly invariant under these transformations we do find that there is strong overlap between the CF-states at momentum ${\bf K}$ and the corresponding particle-hole or inversion transformed reverse flux CF states at momentum ${-\bf K}$.
We also find that the CF states are typically more symmetric under orbital inversion than under particle-hole conjugation.

In \SecRef{sec:Proj_wfn} we test how well the CF trial wave functions approximate the lowest energy states in all centre of mass momentum sectors.
We find that there is generally good overlap between the exact states and the CF states in all sectors, although the best results are obtained in the sectors with the lowest ground state energies and largest gaps. Performing symmetrization under orbital inversion or particle-hole symmetrization typically does not significantly increase the overlap.

In \SecRef{sec:Charged_excitations} we consider charged excitations of the CF liquid which occur when the flux is raised or lowered by one quantum from the flux of the CF liquid ground state. Natural trial wave functions for these states are given by considering CFs subject to a single effective flux, at level $n=N_e$ of the Jain series,  \ie at filling $\nu=\frac{N_e}{2N_e\pm1}$ where there is one CF Landau level per electron. We find that these states indeed provide a good description of the electron system. 

In light of the recent discussion of whether the composite fermions are Dirac fermions, it is interesting to investigate whether the electron system can be modeled better by noninteracting non-relativistic fermions with quadratic dispersion, or by noninteracting Dirac fermions with linear dispersion. We attempt to fit the CF dispersion from the exact energies in \SecRef{sec:linear_vs_quadratic}, but we find that the exact spectrum can be fit well with a wide range of CF dispersions and we are thus not able to resolve the question of linear vs. quadratic dispersion.

In \SecRef{sec:Berry_phase} we attempt to extract the Berry phase which occurs when one CF is dragged around the CF Fermi disc. Here, we see qualitatively similar features to those which were announced for larger systems using approximate methods\cite{Wang2017}, but finite size effects make any definite interpretation problematic.

Further discussion  can be found in~\SecRef{sec:Discussion}.

\section{Composite fermion wave functions}\label{sec:Wave_Func}
\subsection{The CF construction at $\mathbf{\nu=\frac{1}{2}}$}
Electronic wave functions for the fractional quantum Hall effect can often be accurately described by a wave-function of composite fermions. For a system of $N_e$ electrons on the plane, we can write these trial wave functions in the form 
\begin{equation}
\Psi_{\mathrm{trial}}(\mathbf{z})=P_{\mathrm{LLL}}
\left[
\psi_{CF}(\mathbf{z})\left(\prod_{i<j}(z_i-z_j)^{2m}\right) e^{-\frac14\sum_j|z_j|^{2}}\right].
\label{eq:CF_wf}
\end{equation} 
Here, the $z_i$ are complex coordinates on the plane and $P_{\mathrm{LLL}}$ represents projection onto the lowest Landau level of the electron system. The factor $\left(\prod_{i<j}(z_i-z_j)^{2m}\right) e^{-\frac14\sum_j|z_j|^2}$ is the so called flux attachment factor. It is also the Laughlin wave function for a system of bosons at filling $\nu=\frac{1}{2m}$. The remaining factor $\psi_{CF}(\mathbf{z})$ can be thought of as the wave function of an effective system of composite fermions, particles formed from electrons by attaching $2m$ flux quanta. Concretely, $\psi_{CF}$ is usually a Slater determinant, 
\[
\psi_{CF}(\mathbf{z})=\det\left[\psi_k(z_j)\right]
\]
containing $N_e$ orbitals of a system with a reduced magnetic flux, the effective flux seen by the composite fermions. The effective flux is chosen so that the total flux of the trial wave function is the desired flux $N_s$ of the electron system. In the case shown, the flux attachment factor captures $2m(N_e-1)$ flux quanta (the flux corresponds to the highest power of any electron coordinate). Hence the effective flux is $N_s-2m(N_e-1)$.

In the description of gapped quantum Hall states, the effective flux is nonzero and the orbitals in the determinant are organized into Landau levels. However, in the special case where $m=1$ and $N_s=2(N_e-1)$, the effective flux vanishes and the orbitals in the Slater determinant are simply orbitals of free electrons in zero magnetic field on the plane, in other words, plane waves $e^{\i \mathbf k\cdot\mathbf r}$ with well defined two dimensional momentum $\mathbf k$. 
The hope is always that the system can be thought of as non-interacting CFs, that is, the Slater determinant will give a good trial wave function and the choice of orbitals which take part in it can be obtained from some single particle energy model for the CFs -- usually, but not necessarily just the single particle energies of the problem which yields the orbitals.   In the problem with effective flux equal to zero, the orbitals should fill a Fermi disc, minimizing the sum of the single particle CF energies $\epsilon_{\mathbf{k}}$.

\subsection{Specifics and symmetry structure on the torus}
A problem that immediately arises when constructing such a wave function on a plane or disc is that the possible wave vectors $\mathbf{k}=(k_x,k_y)$ form a continuous set. 
Considering the system with periodic boundary conditions (\ie on a torus) naturally makes the number of momentum states finite and allows trial wave functions at finite $N_e$ to be defined properly.  Using the torus also brings some further advantages. For example the state with zero effective flux will now occur at $N_s=2 N_e$, \ie precisely at half filling, which is useful in the investigation of particle-hole duality. Also the absence of a physical boundary simplifies the system considerably, and hopefully makes finite systems more representative of the bulk of the system's thermodynamic limit.  

We choose the torus to be defined by lattice vectors, represented as complex numbers $L_1=L\in\RR$ and $L_2=\tau L$, were $\tau=\tau_1+\i\tau_2$ is a complex parameter encoding the geometry of the torus . We will refer to the cases $\tau=\i$ and  $\tau=\frac12+\i\frac{\sqrt3}2$ as the square and hexagonal torus, respectively.
The toroidal problem is only well defined if the torus is pierced by an integer number of flux quanta which sets the torus area to $A=L^2\tau_2=2\pi N_{s} \ell_B^2$, where $\ell_B$ is the magnetic length.
The CF construction introduced on the plane carries over directly to the torus with the only change that the flux attachment factor in Eq. (\ref{eq:CF_wf}) is replaced by a Laughlin state $\Phi_{\frac{1}{2}}\left(\mathbf{z}\right)$ for the torus\cite{Haldane85}. Since the Laughlin state on the torus is not unique, this leaves some ambiguity and it appears that there may be multiple sets of trial wave functions depending on the chosen Laughlin ground state. However, we show in App.~\ref{sec:Laughlintorus_app} that a change of the choice of Laughlin ground state can be compensated with a global shift of the momenta of the composite fermions.  

For the continued discussion, we need to give some information regarding the structure of momentum sectors.
One may define magnetic translation operators which in Landau gauge take the form
\[ t(a+\i b) = e^{ a\partial_x + b\partial_y + \i 2\pi N_s \frac{xb}{\ell_B^2}} \]
and define periodic boundary conditions such that $t(L)\psi(z)=t(\tau L)\psi(z)=\psi(z)$.
Magnetic translations do not in general commute. In fact, we may define small translations $t_j=t(L_j/N_s)$ that satisfy  $t_1t_2=e^{\i \frac{2\pi}{N_s}}t_2t_1$.
Using  these, we define the center of mass translations
\[
T_{j}=\prod_{k=1}^{N_{e}}t_{j}^{\left(k\right)}=\prod_{k=1}^{N_{e}}t^{\left(k\right)}\left(\frac{L_{j}}{N_{s}}\right)
\]
which satisfy the relation $T_1T_2=e^{\i2\pi\nu}T_2T_1$.
We note that at half filling $(T_2)^2$ commutes with $T_1$ and the Hamiltonian and thus $\{H,T_1,T_2^2\}$ form a good set of commuting operators, with associated well defined quantum numbers.
As $T_j^{N_s}=1$ we define the $K_1$ and $K_2$ quantum numbers of a state as
\begin{eqnarray*}
T_{1}\psi & = & e^{\i2\pi\frac{K_{1}}{N_{s}}}\psi\\
(T_{2})^{2}\psi & = & e^{\i2\pi\frac{2K_{2}}{N_{s}}}\psi = e^{\i2\pi\frac{K_{2}}{N_{e}}}\psi.
\end{eqnarray*}
The bosonic wave function  $\Phi_{\nu=\frac{1}{2}}\left(z\right)$ can be chosen such that it has $(K_1,K_2)=(0,0)$ (see App.~\ref{sec:Laughlintorus_app}).
Using this choice we find that the eigenvalues for $\PsiCFL$ from \eqref{eq:CF_wf} under a full revolution of the $j^{\rm th}$ particle around the torus handles are
\begin{eqnarray*}
t^{\left(j\right)}\left(L\right)\PsiCFL & = & e^{\i Lk_{x}^{(j)}}\PsiCFL\\
t^{\left(j\right)}\left(\tau L\right)\PsiCFL & = & e^{\i L\left(\tau_{1}k_{x}^{(j)}+\tau_{2}k_{y}^{(j)}\right)}\PsiCFL.
\end{eqnarray*}
Requiring periodic boundary conditions restricts the set of momenta and we have 
\begin{equation}
  k=k_x+\i k_y=k_1G_1+k_2G_2,\label{eq:k}
\end{equation}
where the reciprocal lattice vectors are $G_1 =-\frac{\i2\pi}{L\tau_{2}}\tau$ and $G_2 =\i\frac{2\pi}{L\tau_{2}}$.
This dual lattice has geometry $\tau_G=\frac{G_{2}}{G_{1}}=-\frac{1}{\tau}$ which is related to the direct lattice by a modular $S$-transform.
We will abuse the notation a bit here and alternatively let $\mathbf k$ refer to the physical momentum $(k_x,k_y)$ or the reciprocal lattice indices $(k_1,k_2)$.
This abuse of notation extends to the total momentum $\mathbf K=\sum_j\mathbf k^{j}$.

In using the indices $k_1$ and $k_2$ the many body quantum numbers of the CF-FL state are then
\begin{eqnarray*}
T_1\PsiCFL & = & e^{\i\frac{2\pi}{N_{s}}\sum_{j=1}^{N_{e}}k_{1}^{j}}\PsiCFL\\
T_2^2\PsiCFL & = & e^{\i\frac{2\pi}{N_{e}}\sum_{j=1}^{N_{e}}k_{2}^{j}}\PsiCFL
\end{eqnarray*}
giving $K_{1}=\sum_{j=1}^{N_{e}}k_{1}^{(j)}$ and $K_{2}=\sum_{j=1}^{N_{e}}k_{2}^{(j)}$.

We here see that we can write a wave function for any desired momentum $(K_1,K_2)$ by choosing an appropriate set $\{(k_1,k_2)\}$ to use in the plane wave factor.

\subsection{Lowest Landau level projection}
At half filling on the torus we may perform the lowest Landau level projection which appears in \eqref{eq:CF_wf} exactly. Normally this is not straight forward on a torus, but here it is facilitated by the absence of fluxes in the plane wave factor.  The projected state is given by 
\begin{equation}
\PsiCFL=e^{-\frac{1}{2}\sum_{j}\left|\mathbf{k}^{(j)}\right|^{2}}\det\left(t^{(i)}_{-k^{(j)}_2,k^{(j)}_{1}}\right)\psi_{\frac{1}{2}}
\label{eq:CF_projected}
\end{equation}
where $t_{n,m}=t\left(\frac1{N_{s}}(nL_1+mL_2)\right)$ is a finite translation on the $N_s\times N_s$ grid of boundary condition preserving translations.
Note that the direction of the translations is perpendicular to the direction of the momenta $\mathbf k^{(j)}$ since $\vec L_m\cdot \vec G_n =\delta_{mn}$.
Note also that the above result is a gauge invariant statement and not an artifact of the choice of Landau gauge. A detailed derivation of \eqref{eq:CF_projected} is given in App.~\ref{sec:LLL_proj}.

We define an FS (Fermi surface) configuration to be a single Slater determinant of plane wave states.
We note that the effect of multiplying with a plane wave before projection is
the same (up to a scale factor) as adding a translation operator after projection.
From this result we draw two non-trivial conclusions.
The first is that changing $k_1\to k_1 + N_s$ will not change the actual state after projection, only the normalization.
Secondly, changing all $k_{1}^{(j)}\to k_{1}^{(j)}+1$ is the same as adding a global translation $T_2$ after projection. This does not change the state after projection in any qualitative way except to transform the state at $K_1$ into its degenerate copy at $K_1+N_e$. Similar statements hold for $k_2$ and $T_2$.

While  the exact projection is given by (\ref{eq:CF_projected}), this expression does not yield an efficient algorithm for evaluation of the wave functions, because of the need to anti-symmetrize over translation operators. In all our numerical work in the rest of this paper we therefore make use of the energy projection which we introduced in Ref.~\onlinecite{Fremling_16}. This allows for essentially exact projection of any reasonable trial wave function at system sizes close to those accessible to exact diagonalization. 
To deal with trial wave functions for larger system sizes, one may consider using approximate projection schemes or even modified trial wave functions which are not fully in the LLL. References \onlinecite{Shao2015} and \onlinecite{Wang2017} represent recent advances in these directions. Since we are mostly interested in comparison with exact wave functions here, we do not pursue these approximate methods.

\section{Energy landscapes. }\label{sec:Landscapes}
There are vastly more distinguishable FS configurations (choices of $\mathbf{k}^{1},...,\mathbf{k}^{N_{e}}$) than actual energy eigenstates at a particular momentum $\mathbf{K}$. (There are $\binom{N_s}{N}$ states in the LLL and $\binom{N_{s}^{2}}{N}$ FS configurations.)
It is therefore important to find an estimate for the variational energy that a FS state will have after projection,
in order to select a trial space in which to construct composite fermions.

\begin{figure}[tb]
 \begin{tabular}[t]{ccc}
   &Composite fermion energy & Exact Ground state energy
   \tabularnewline
   \begin{turn}{90}\quad\quad\quad\quad\quad Square\end{turn}&
  \includegraphics[width=0.45\linewidth]{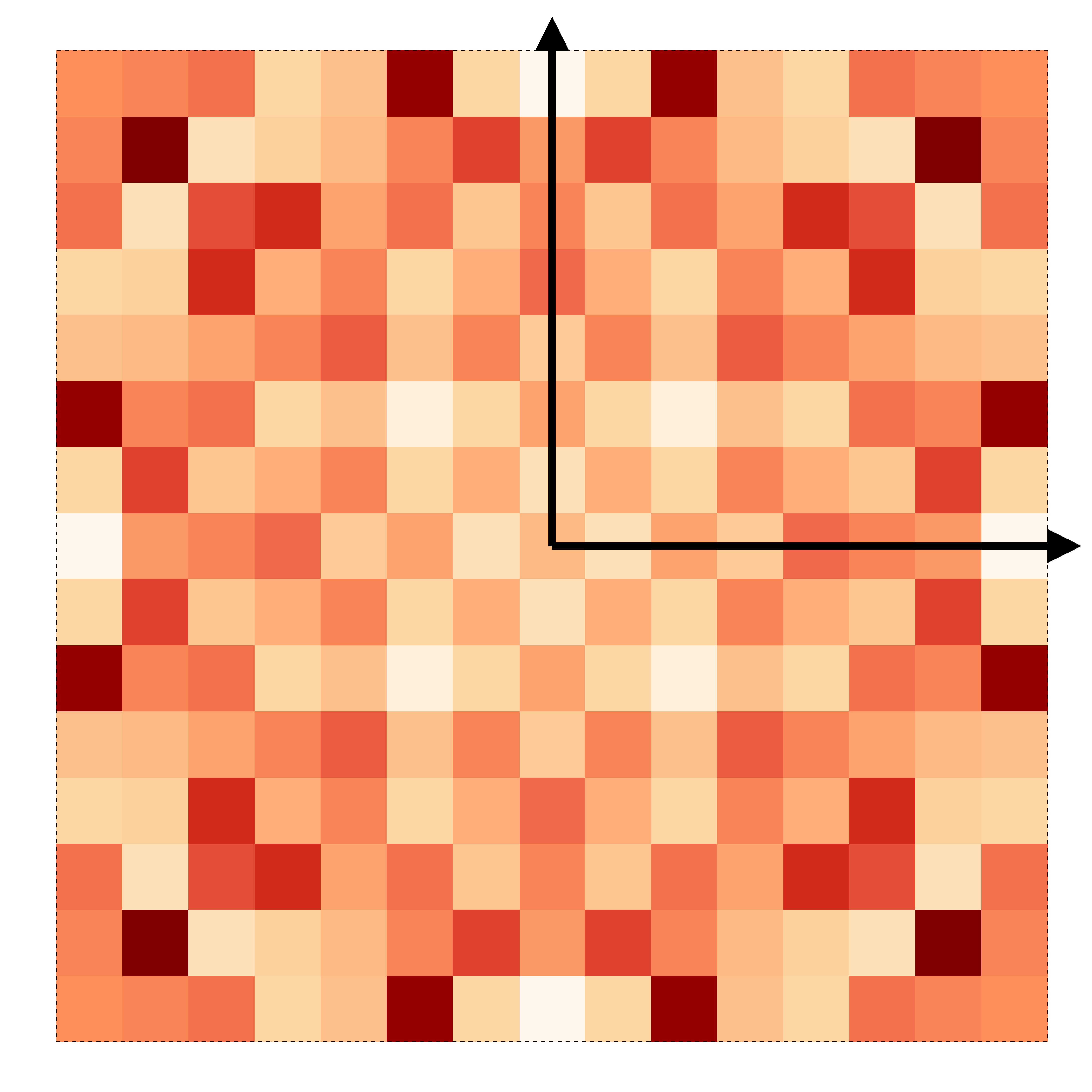}&
  \includegraphics[width=0.45\linewidth]{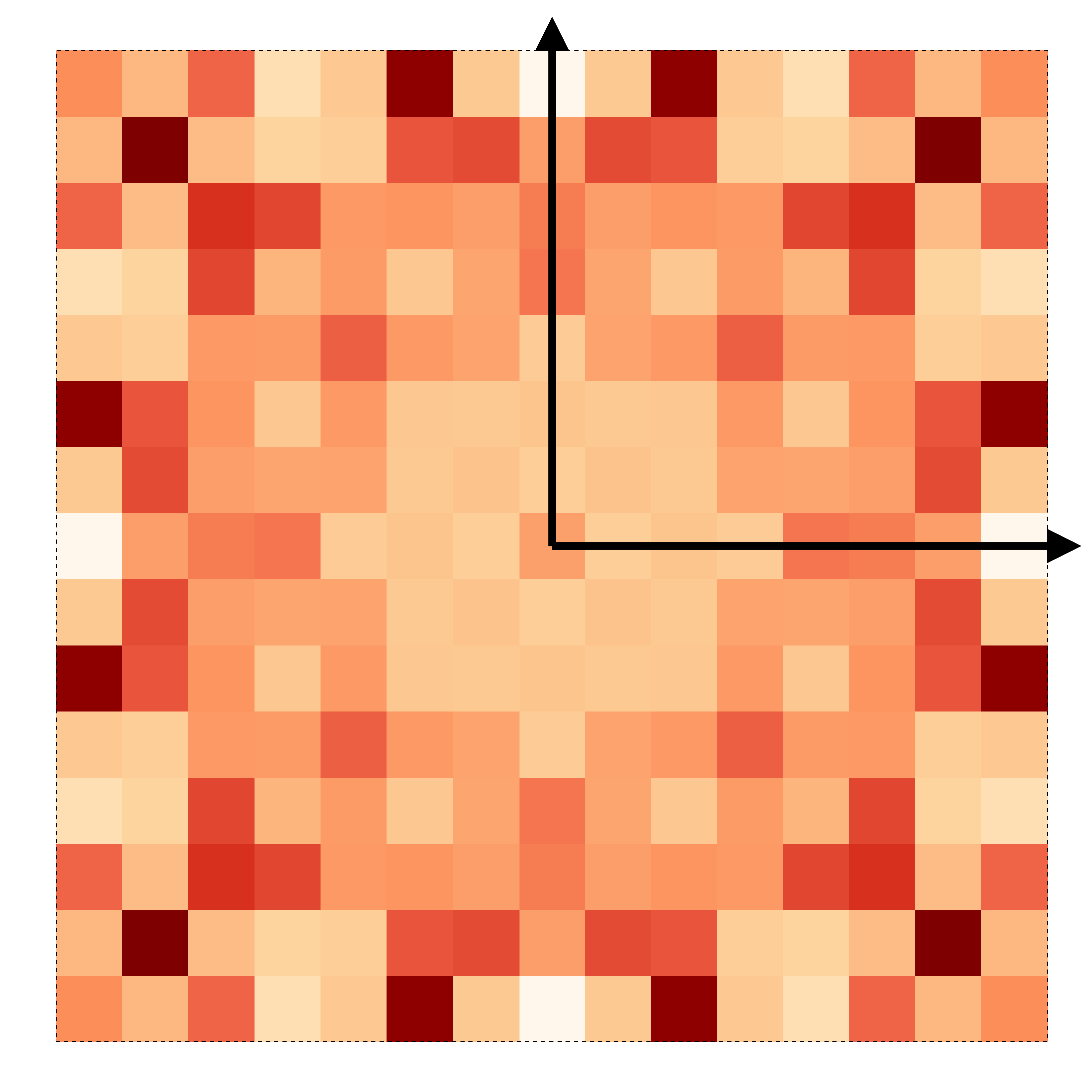}
  \tabularnewline
  \begin{turn}{90}\quad\quad\quad\quad\quad Hexagon\end{turn}&
  \includegraphics[width=0.45\linewidth]{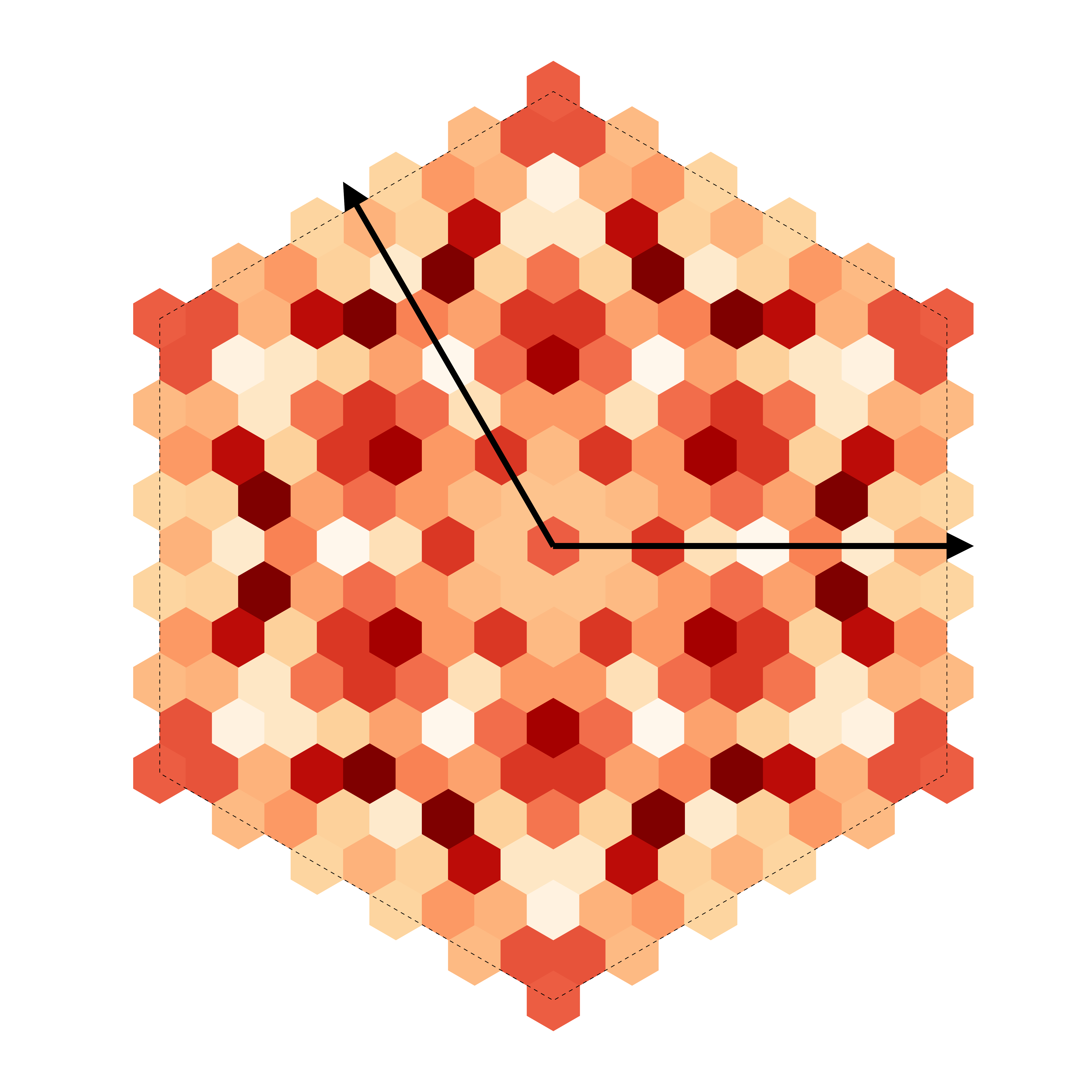}&
  \includegraphics[width=0.45\linewidth]{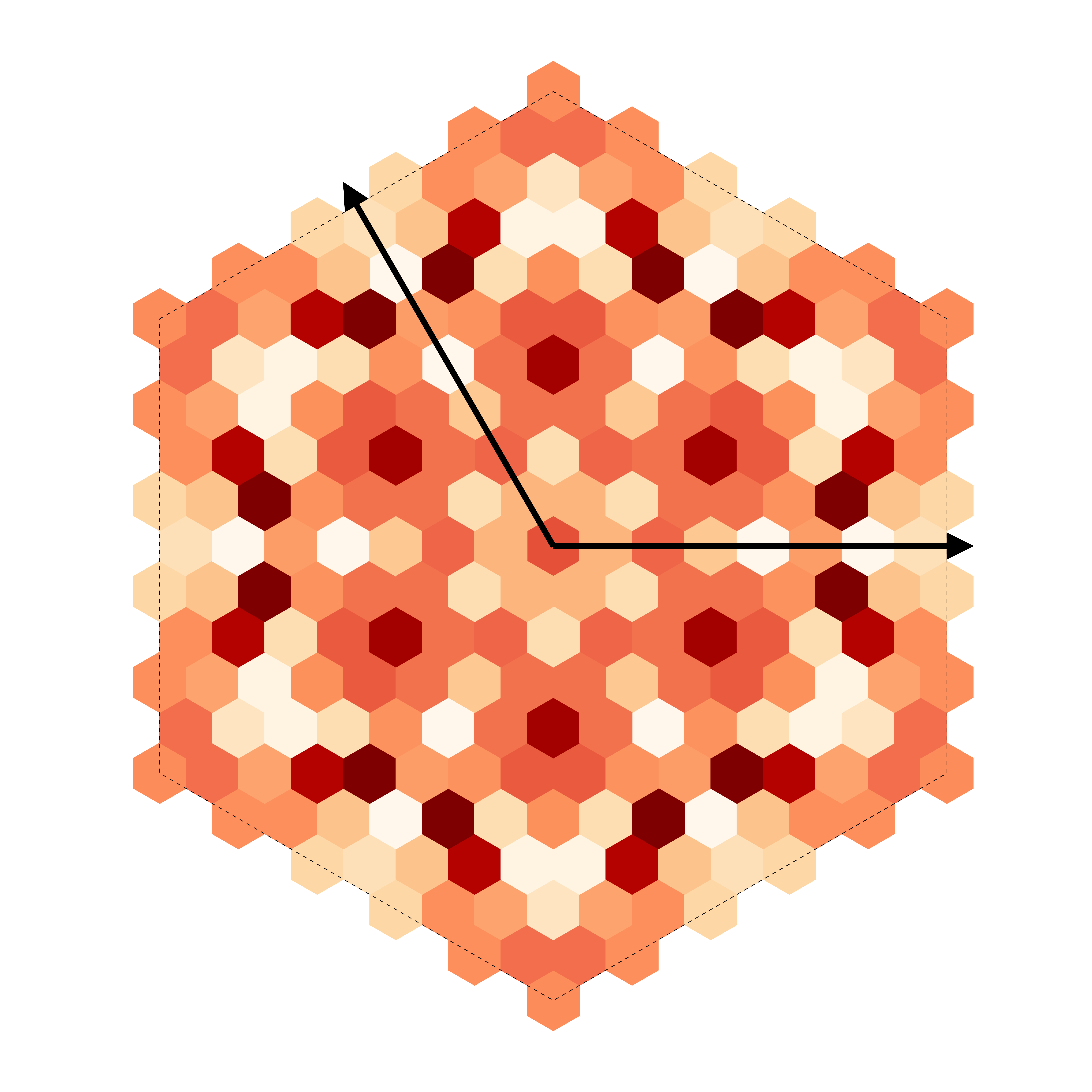}
  \end{tabular}

\caption{Comparison between exact energy landscape and composite fermion energy
landscape, given by \eqref{eq:cf_energy} for $N_e=15$. The lowest (highest) energy is at the darkest (lightest) red. The ground states at $(K_1,K_2)=(6,6)$ and $(K_1,K_2)=(5,6)$ respectively are accurately identified by the CF model, as well as the low energy features. The arrows show the directions of the two reciprocal unit vectors $\mathbf{G}$. Energy is in arbitrary units.\label{fig:energy_landscape}}
\end{figure}

\begin{table}
  \begin{tabular}{c|cr|cr} 
    $N_e$ & Square & deg. & Hexagonal & deg.
    \tabularnewline
    \hline
    7 & (0,2) & 4 & (0,0) & 1\\
    8 & (1,1) & 4 & (1,2) & 6\\
    9 & (0,0) & 1 & (0,3) & 6\\
    10 & (0,2) & 2 & (0,5) & 6\\
    11 & (4,5) & 8 & (2,-4) & 12\\
    12 & (6,6) & 4 & (4,-4) & 6\\
    13 & (5,5) & 4 & (5,-6) & 12\\
    14 & (4,7) & 8 & (0,7) & 6\\
    15 & (6,6) & 4 & (5,6) & 12\\
    \tabularnewline
  \end{tabular}
  \caption{Momentum sectors ($K_1$,$K_2$) containing the globally minimal total CF-energy $E_\mathrm{CF}$ for square and hexagonal torus. The number of momentum sectors with the same minimal $E_\mathrm{CF}$  is listed as 'deg.'.\label{tab:GS-states}}
\end{table}

\begin{figure}
\begin{tabular}{c}

  \begin{tikzpicture}
    \coordinate (CM) at (6/15,6/15);
    \coordinate (E1) at (0,0);
    \coordinate (E2) at (0,1);
    \coordinate (E3) at (1,0);
    \coordinate (E4) at (1,1);
    \coordinate (E5) at (-1,0);
    \coordinate (E6) at (-1,1);
    \coordinate (E7) at (0,-1);
    \coordinate (E8) at (0,2);
    \coordinate (E9) at (1,-1);
    \coordinate (E10) at (1,2);
    \coordinate (E11) at (-1,-1);
    \coordinate (E12) at (-1,2);
    \coordinate (E13) at (2,0);
    \coordinate (E14) at (2,1);
    \coordinate (E15) at (2,-1);
    
    \foreach \x in {-2,...,2}
             {\draw[dashed] (\x,-2) --  (\x,2);
               \draw[dashed] (-2,\x) --  (2,\x);}
             \draw[line width=3pt,->] (0,-2) --  (0,3) node[above]{$G_2$};
             \draw[line width=3pt,->] (-2,0) --  (3,0) node[right]{$G_1$};
    \draw[fill=red] (E1) circle (.25) ;
    \draw[fill=red] (E2) circle (.25) ;
    \draw[fill=red] (E3) circle (.25) ;
    \draw[fill=red] (E4) circle (.25) ;
    \draw[fill=red] (E5) circle (.25) ;
    \draw[fill=red] (E6) circle (.25) ;
    \draw[fill=red] (E7) circle (.25) ;
    \draw[fill=red] (E8) circle (.25) ;
    \draw[fill=red] (E9) circle (.25) ;
    \draw[fill=red] (E10) circle (.25) ;
    \draw[fill=red] (E11) circle (.25) ;
    \draw[fill=red] (E12) circle (.25) ;
    \draw[fill=red] (E13) circle (.25) ;
    \draw[fill=red] (E14) circle (.25) ;
    \draw[fill=red] (E15) circle (.25) ;
    \draw[fill=blue] (CM) circle (.15) ;
  \end{tikzpicture}

  \\
  
    \begin{tikzpicture}
    \coordinate (G1) at (0.866,-.500);
    \coordinate (G2) at (0,1.000);
    \coordinate (CM) at ($5/15*(G1)+6/15*(G2)$);
    \coordinate (E1) at ($0*(G1)-0*(G2)$);
    \coordinate (E2) at ($1*(G1)+1*(G2)$);
    \coordinate (E3) at ($0*(G1)+1*(G2)$);
    \coordinate (E4) at ($1*(G1)-0*(G2)$);
    \coordinate (E5) at ($-1*(G1)-0*(G2)$);
    \coordinate (E6) at ($0*(G1)-1*(G2)$);
    \coordinate (E7) at ($-1*(G1)-1*(G2)$);
    \coordinate (E8) at ($1*(G1)+2*(G2)$);
    \coordinate (E9) at ($2*(G1)+1*(G2)$);
    \coordinate (E10) at ($2*(G1)+2*(G2)$);
    \coordinate (E11) at ($-1*(G1)+1*(G2)$);
    \coordinate (E12) at ($0*(G1)+2*(G2)$);
    \coordinate (E13) at ($1*(G1)-1*(G2)$);
    \coordinate (E14) at ($2*(G1)-0*(G2)$);
    \coordinate (E15) at ($-2*(G1)-1*(G2)$);

    \draw[dashed] ($-2*(G1)-2*(G2)$) --  ($0*(G1)-2*(G2)$);
    \draw[dashed] ($-2*(G1)-1*(G2)$) --  ($1*(G1)-1*(G2)$);
    \draw[dashed] ($-2*(G1)-0*(G2)$) --  ($2*(G1)-0*(G2)$);
    \draw[dashed] ($-1*(G1)+1*(G2)$) --  ($2*(G1)+1*(G2)$);
    \draw[dashed] ($-0*(G1)+2*(G2)$) --  ($2*(G1)+2*(G2)$);

    \draw[dashed] ($-2*(G1)-2*(G2)$) --  ($-2*(G1)+0*(G2)$);
    \draw[dashed] ($-1*(G1)-2*(G2)$) --  ($-1*(G1)+1*(G2)$);
    \draw[dashed] ($-0*(G1)-2*(G2)$) --  ($0*(G1)+2*(G2)$);
    \draw[dashed] ($+1*(G1)-1*(G2)$) --  ($+1*(G1)+2*(G2)$);
    \draw[dashed] ($+2*(G1)-0*(G2)$) --  ($+2*(G1)+2*(G2)$);

    \draw[dashed] ($0*(G1)-0*(G2)-2*(G1)$) --  ($+2*(G1)+2*(G2)-2*(G1)$);
    \draw[dashed] ($-1*(G1)-1*(G2)-1*(G1)$) --  ($+2*(G1)+2*(G2)-1*(G1)$);
    \draw[dashed] ($-2*(G1)-2*(G2)+0*(G1)$) --  ($+2*(G1)+2*(G2)+0*(G1)$);
    \draw[dashed] ($-2*(G1)-2*(G2)+1*(G1)$) --  ($+1*(G1)+1*(G2)+1*(G1)$);
    \draw[dashed] ($-2*(G1)-2*(G2)+2*(G1)$) --  ($0*(G1)+0*(G2)+2*(G1)$);
    
             \draw[line width=3pt,->] ($-2*(G1)$) --  ($3*(G1)$) node[right]{$G_1$};
             \draw[line width=3pt,->] ($-2*(G2)$) --  ($3*(G2)$) node[above]{$G_2$};

    \draw[fill=red] (E1) circle (.25) ;
    \draw[fill=red] (E2) circle (.25) ;
    \draw[fill=red] (E3) circle (.25) ;
    \draw[fill=red] (E4) circle (.25) ;
    \draw[fill=red] (E5) circle (.25) ;
    \draw[fill=red] (E6) circle (.25) ;
    \draw[fill=red] (E7) circle (.25) ;
    \draw[fill=red] (E8) circle (.25) ;
    \draw[fill=red] (E9) circle (.25) ;
    \draw[fill=red] (E10) circle (.25) ;
    \draw[fill=red] (E11) circle (.25) ;
    \draw[fill=red] (E12) circle (.25) ;
    \draw[fill=red] (E13) circle (.25) ;
    \draw[fill=red] (E14) circle (.25) ;
    \draw[fill=red] (E15) circle (.25) ;
    \draw[fill=blue] (CM) circle (.15) ;
  \end{tikzpicture}

\end{tabular}

\caption{ The FS configurations with the lowest CF-energy for the global ground state at $N_e=15$ for a square and hexagonal torus. These are located at $(K_1,K_2)=(6,6)$ and $(K_1,K_2)=(5,6)$ respectively. The center of mass is at the blue small dot.\label{fig:Good_FS_States}}
\end{figure}
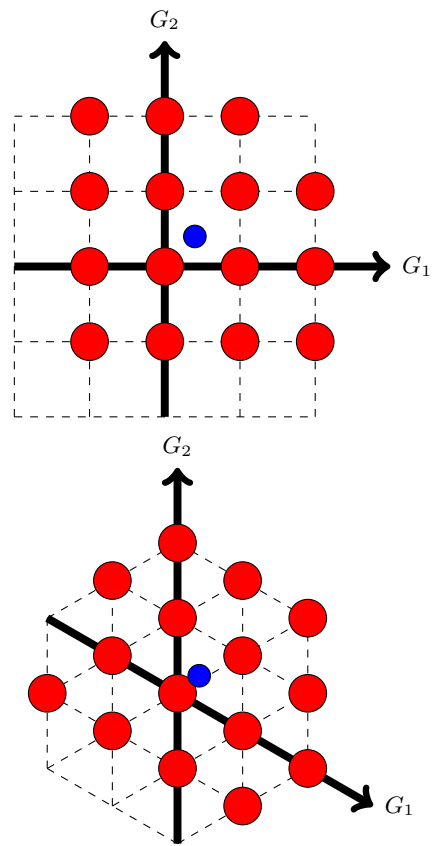

\begin{figure*}[htb]
 \begin{tabular}[t]{c|cc||cc}
    &(5,5) square & (2,3) square & (5,5) hexagon & (2,3) hexagon
    \\
    \hline
    \begin{turn}{90}PH-conjugation\end{turn}&
    \includegraphics[width=.22\linewidth]{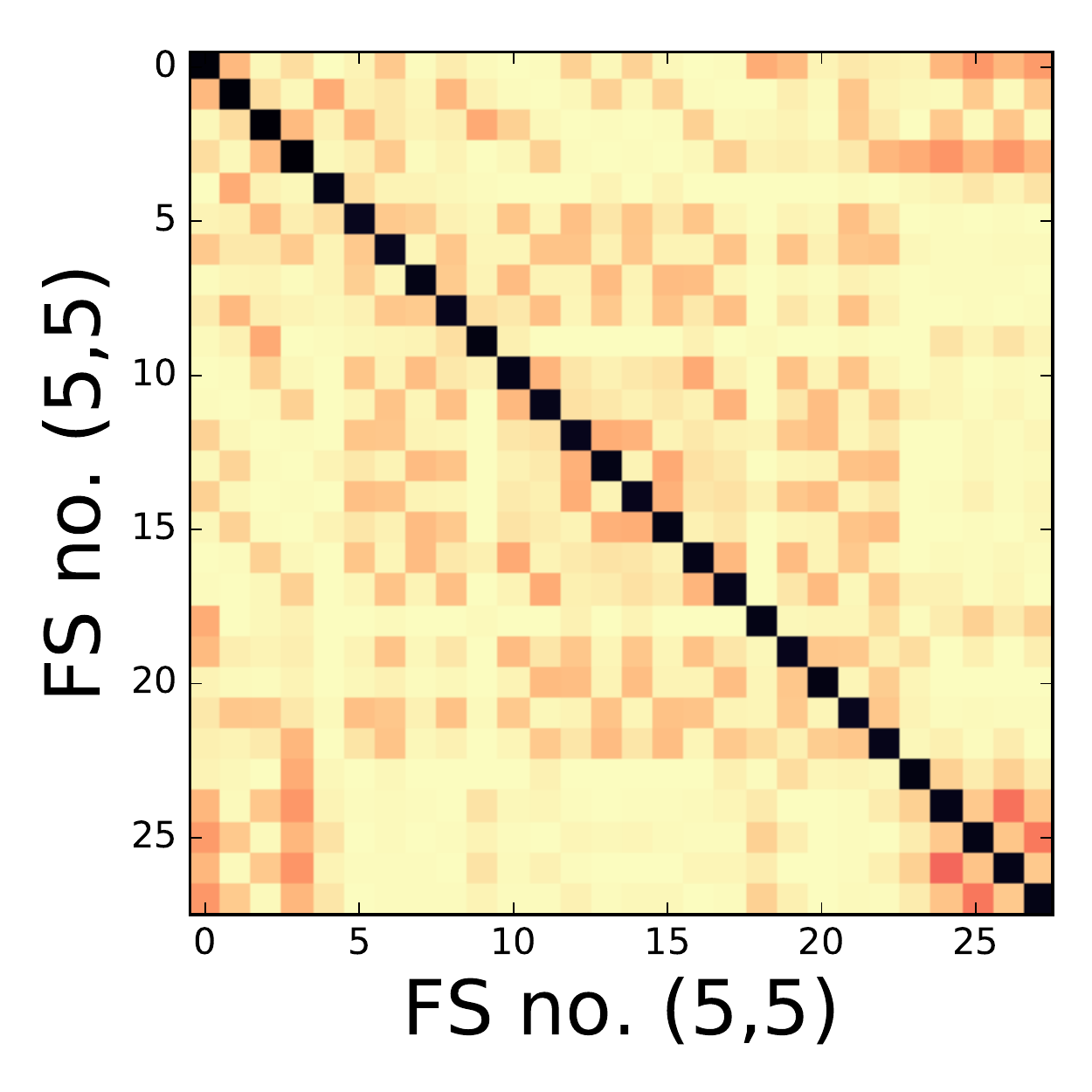}&
    \includegraphics[width=.22\linewidth]{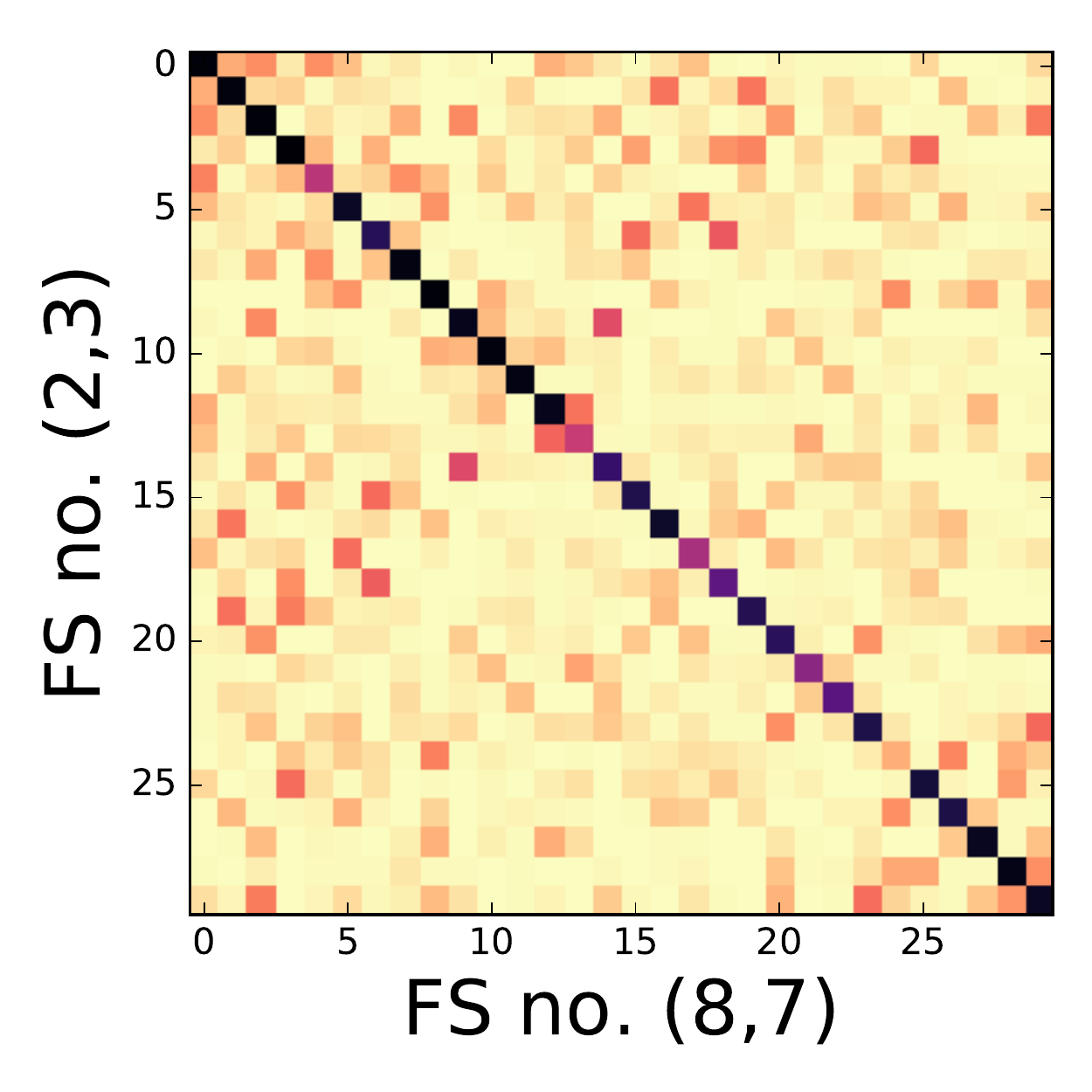}&
    \includegraphics[width=.22\linewidth]{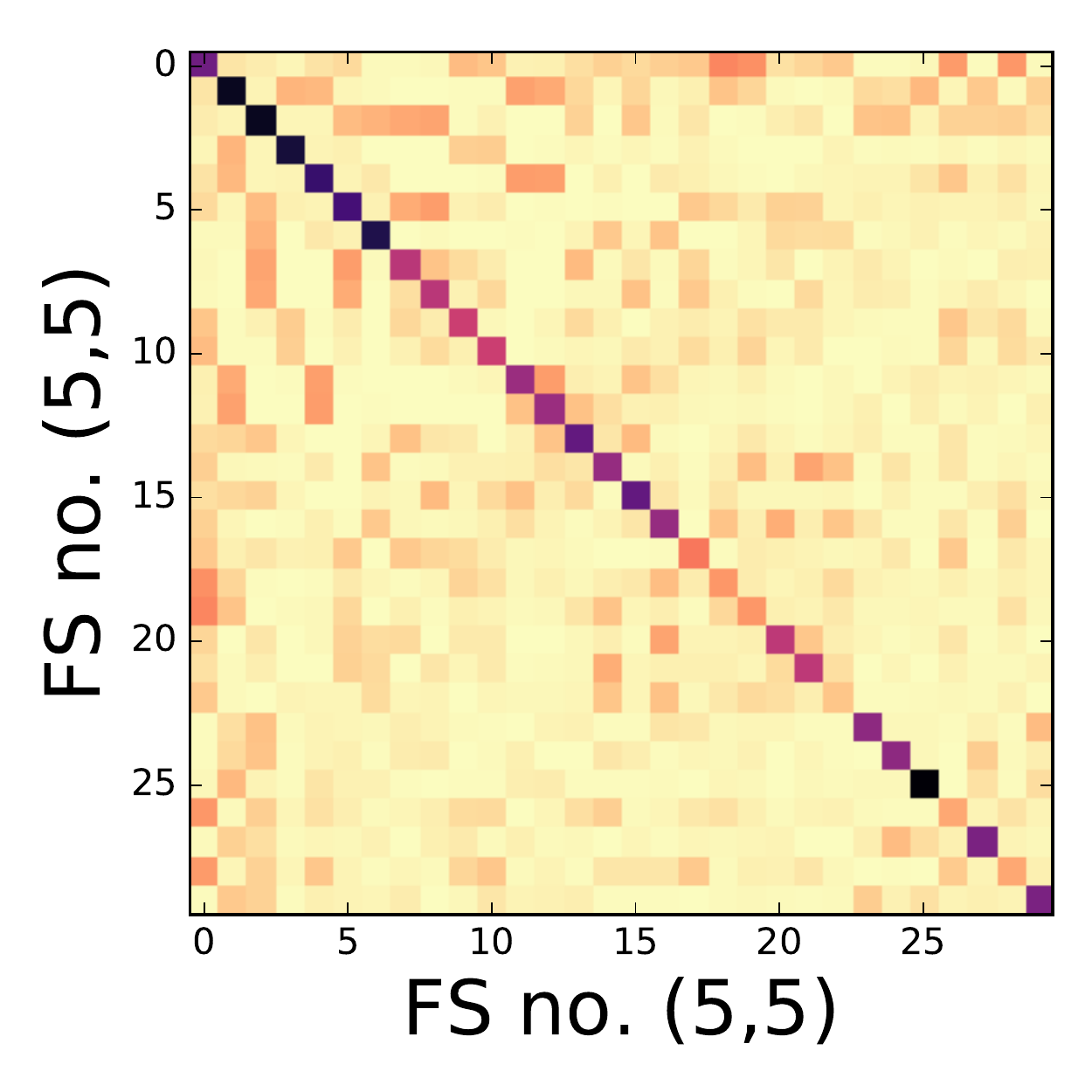}&
    \includegraphics[width=.22\linewidth]{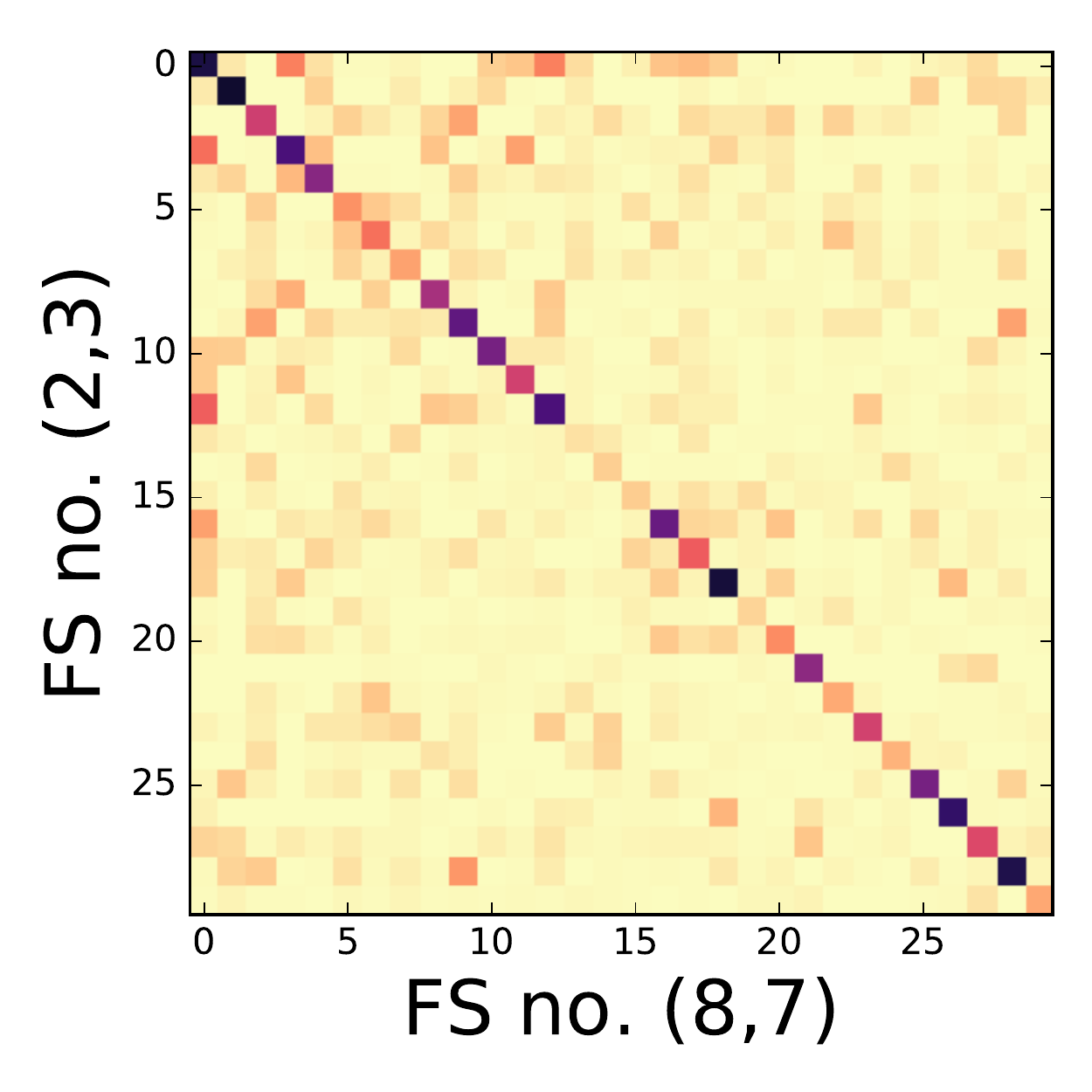}
    \\
    \hline
    \begin{turn}{90}Inversion\end{turn}&
    \includegraphics[width=.22\linewidth]{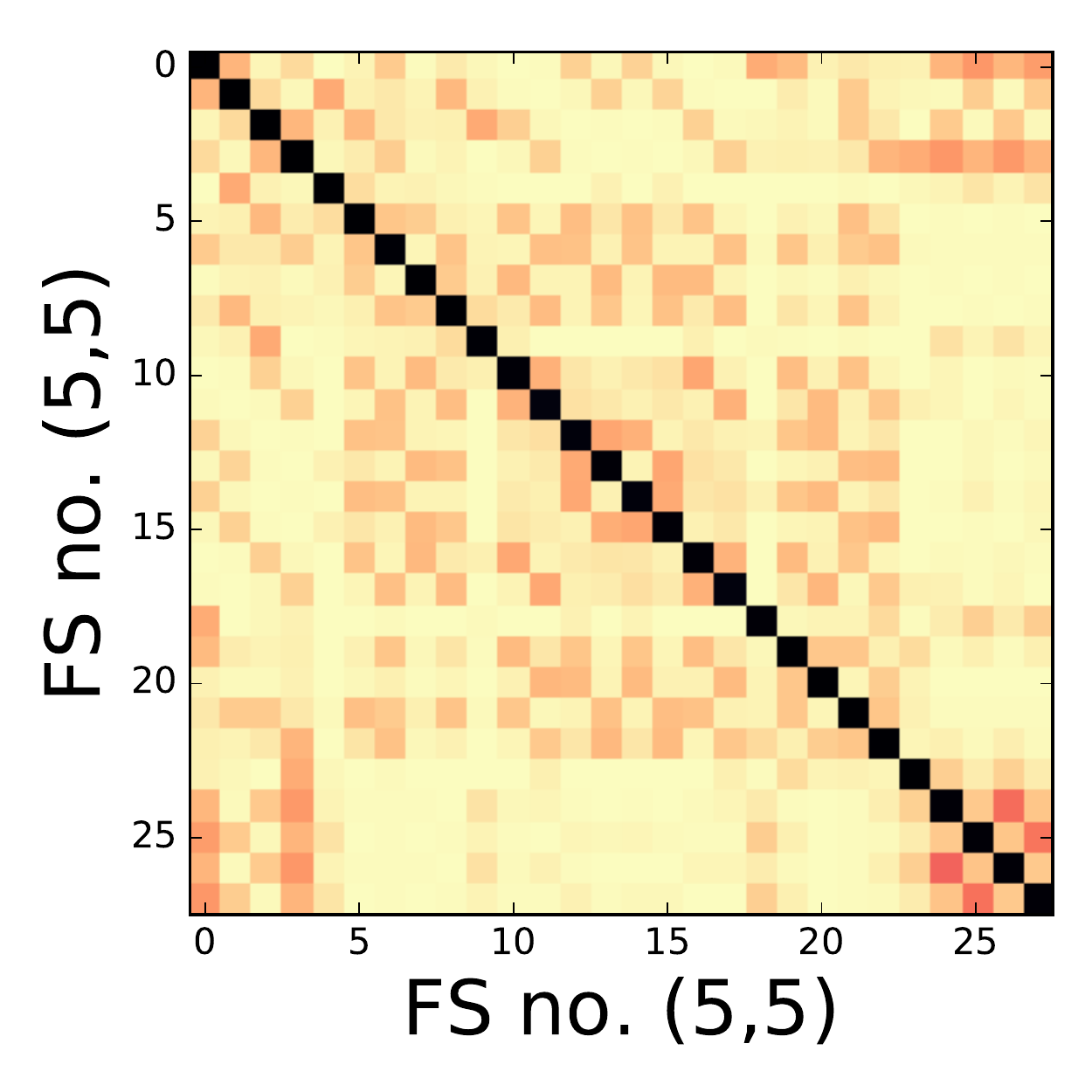}&
    \includegraphics[width=.22\linewidth]{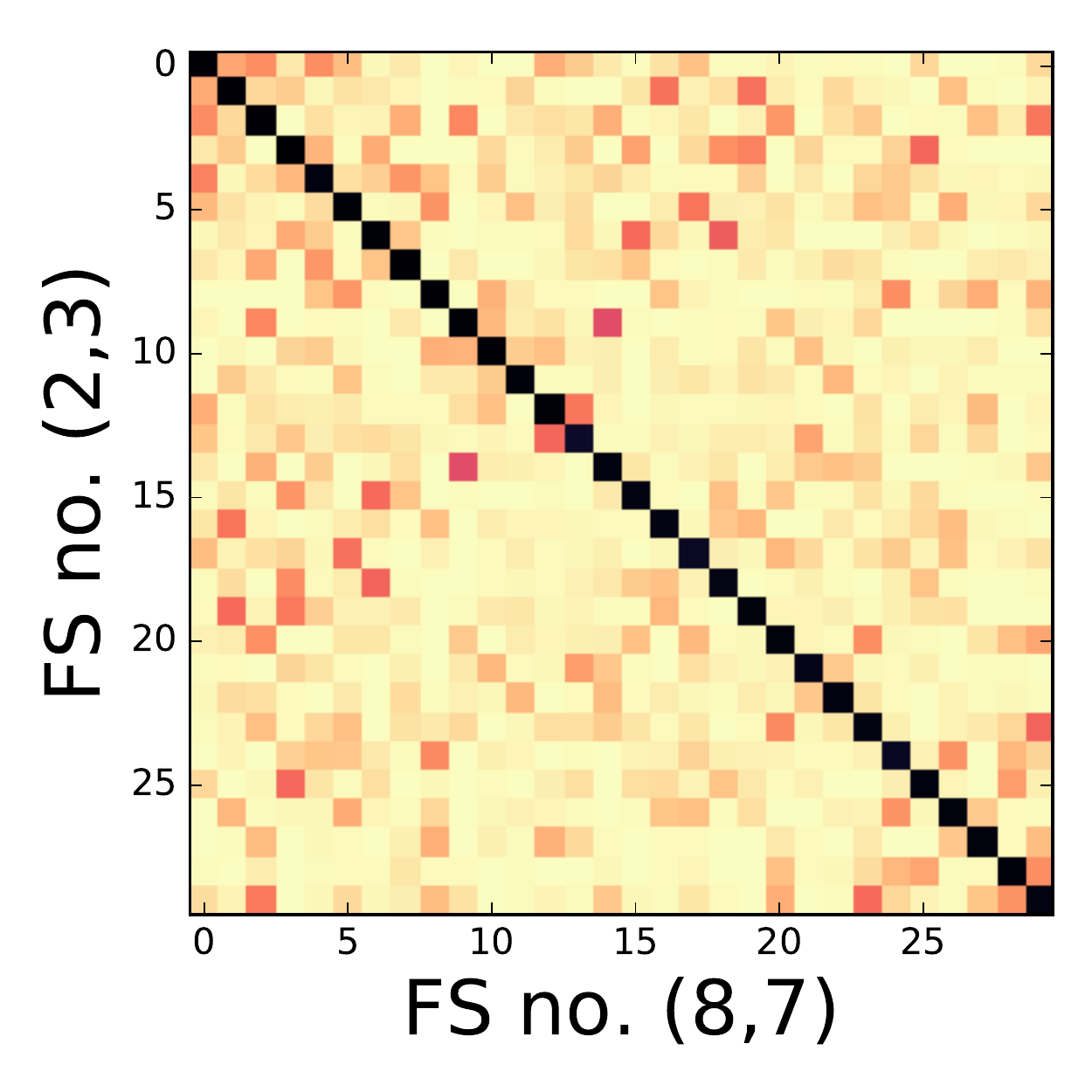}&
    \includegraphics[width=.22\linewidth]{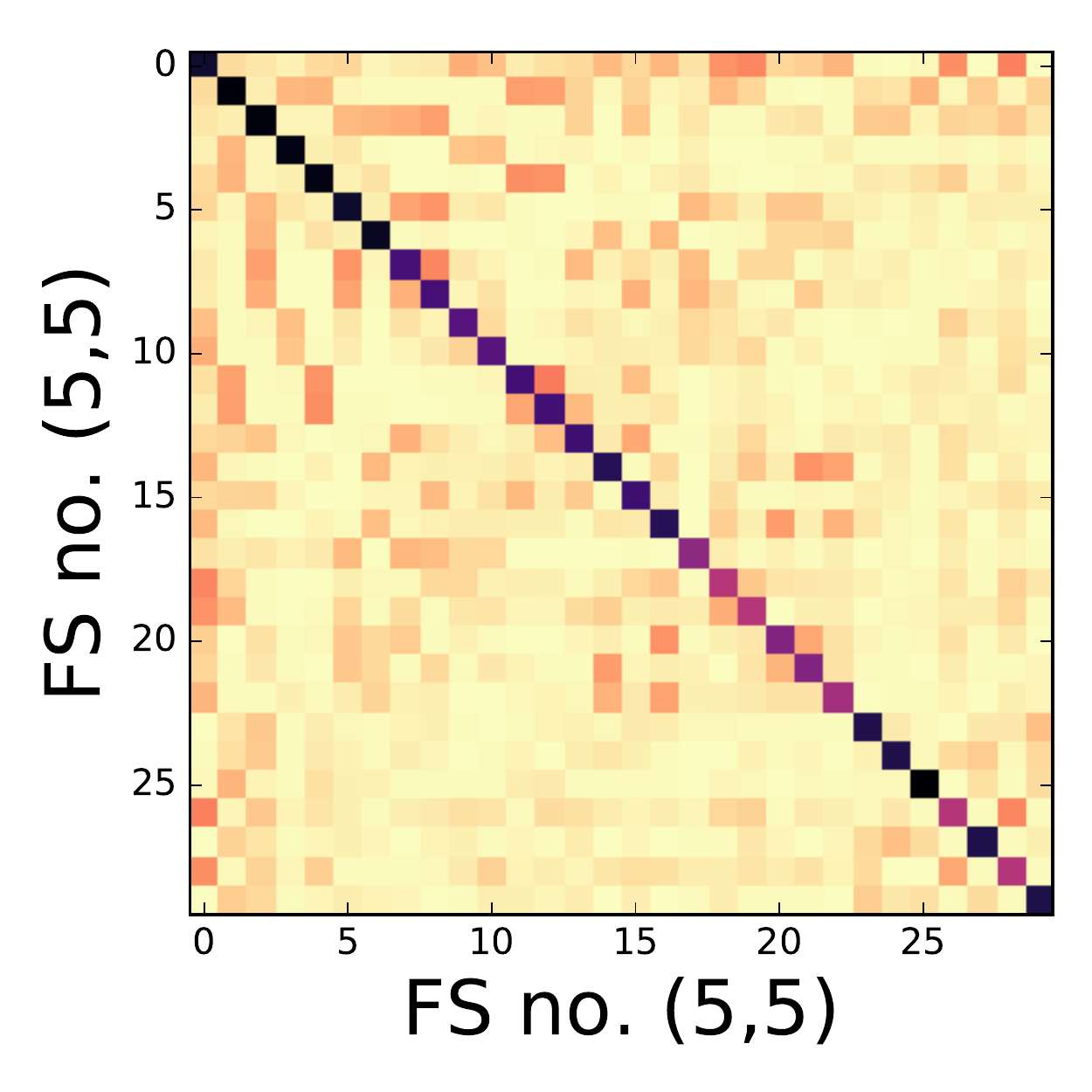}&
    \includegraphics[width=.22\linewidth]{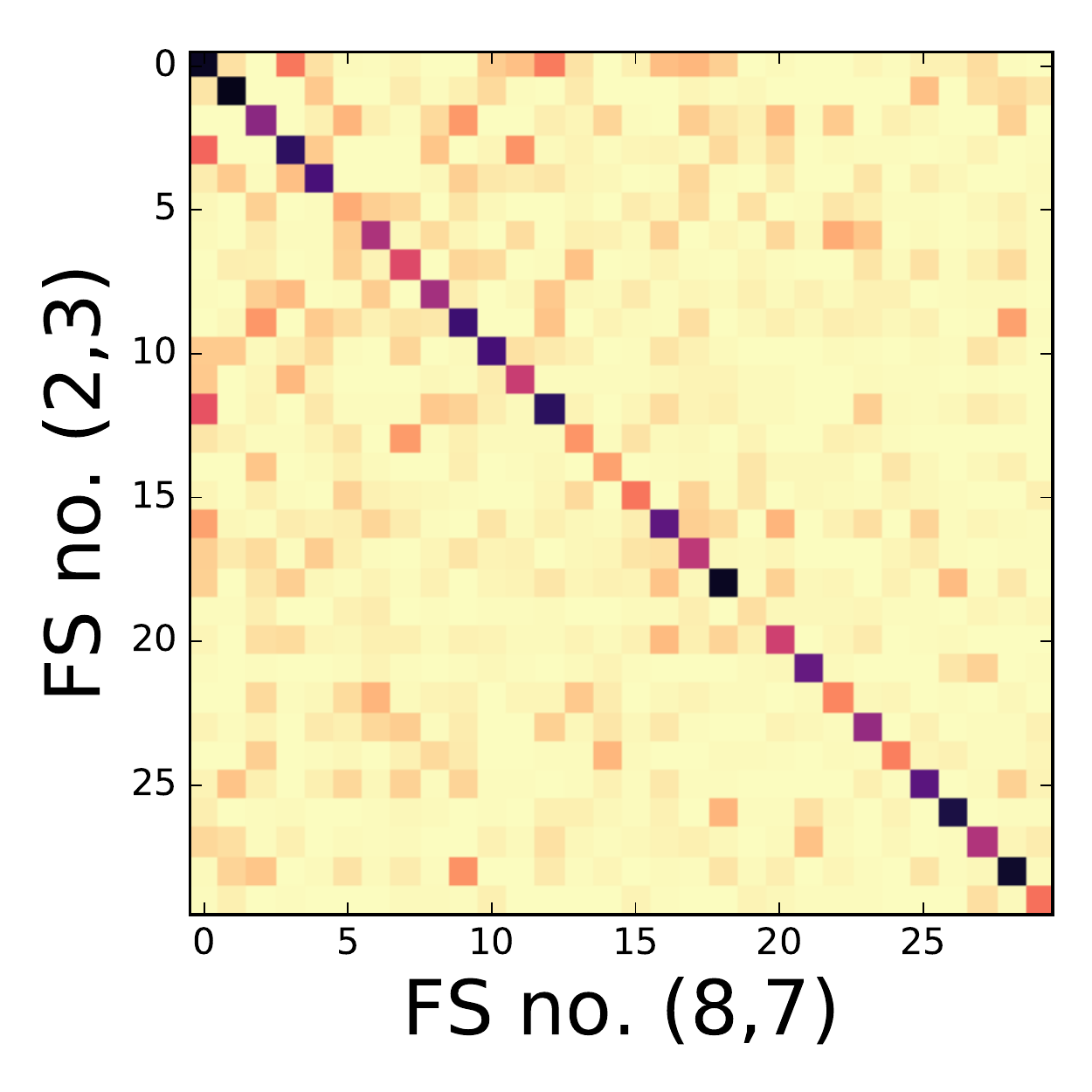}
    \\
    \hline
    & \multicolumn{4}{c}{\includegraphics[width=.80\linewidth]{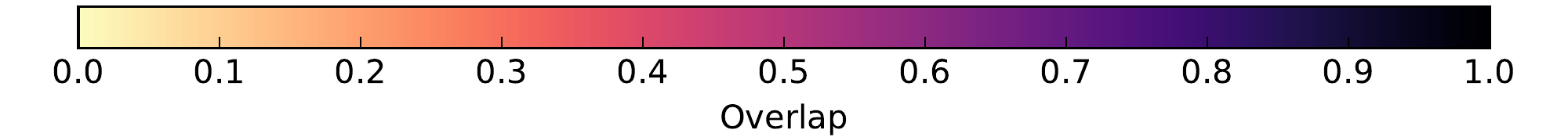}
    }
    \tabularnewline
  \end{tabular}

 \caption{Plot of the overlap between all of the 30 FS-states with smallest CF energy and their PH conjugated/orbital inverted duals for $N_e=10$ in the momentum sectors $(5,5)$ and $(2,3)$.
   Left = Square torus , Right = Hexagonal torus. Upper=Particle hole conjugation, Lower= Orbital inversion.
   The momentum sectors $(2,3)$ and $(8,7)$ are related under PH conjugation/orbital inversion as $(K_1,K_2)\to(N_e-K_1,N_e-K_2)$. At $N_e=10$ the $(5,5)$ sector is self dual.
   As can be seen, in square geometry the overlaps are extremely high with both the inverted state and the PH conjugated state.
   In the hexagonal geometry, only the FS configurations with the lowest CF-energy have high overlap with their inverted and PH conjugated duals.
   The FS-conjurations are in general more inversion symmetric than PH symmetric.
   \label{fig:PH_vs_Inversion_invariance}}
\end{figure*}

For this purpose we assume the composite fermions are non-interacting
and that their single particle energy is given by\cite{Shao2015,Haldane_Unp}
\begin{equation}
\epsilon_{\mathrm{CF}}\left(\mathbf{k}\right)=\left|\mathbf{k}-\frac{\mathbf{K}}{N_{e}}\right|^{d}
\label{eq:cf_energy}
\end{equation}
 where $\mathbf{k}=\left(k_{x},k_{y}\right)$, $\mathbf{K}=\sum_{j=1}^{N_{e}}\mathbf{k}^{(j)}$
 is the total momentum of all the composite fermions and $d$ gives the type of dispersion relation.
 The total CF energy would then be the sum $E_{\mathrm{CF}}=\sum_{j=1}^{N_e}\epsilon_{CF}\left(\mathbf{k}^{(j)}\right)$.
 
 For $d=2$ there is often high degeneracy among the CF states, because there are many different ways to write an integer as a sum of many squares. We can split this by adding a ``surface tension'' term in the energy. This counts the empty orbitals next to the occupied orbitals, and sums over all occupied orbitals. This can be recast formally as $\sum_{\langle i,j\rangle}(N_i(N_j-1)+N_j(N_i-1))$, where $\langle i,j\rangle$ is a sum over nearest neighbour CF orbitals.

There is an ongoing discussion on whether CFs are conventional non-relativistic
fermions or more Dirac like, which at a naive level would translate into choosing between $d=2$ or $d=1$ respectively. 
In this paper we will be using $d=2$, unless otherwise specified.

The offset $\frac{\mathbf{K}}{N_{e}}$ is a peculiarity for composite fermions,
but can be easily argued for by noting that the actual variational energy of any FS state will be invariant under a global reciprocal translation $\mathbf{k}^{(j)}\to\mathbf{k}^{(j)}+\mathbf{G}$.
\footnote{In fact the energy is invariant under any shift $\mathbf k$ but if $\mathbf k$ is not a reciprocal lattice vector the resulting wave function will no longer have periodic boundary conditions.}
In the discussion bellow, we will always choose $\mathbf{k}^{(j)}$ such that it is in the first Brillouin zone.
We reassuringly note that if the shift is not included in \eqref{eq:cf_energy}, then the energy landscape from exact diagonalization is not reproduced even qualitatively.

We note that the CF energy in \eqref{eq:cf_energy} has good qualitative agreement with the exact ground state energy obtained from exact diagonalization, see Fig.~\ref{fig:energy_landscape}.
It gives both a good estimate of the location of the sectors with high energy as well as low energy.
Specifically  \eqref{eq:cf_energy} manages to pinpoint the precise momentum sector of the total ground state as well as the low lying excitations.
We note that the actual variational energy of the FS states in \eqref{fig:energy_landscape} looks similar to the CF-energy/exact energy but we don't show it here.
See however Fig.~\ref{fig:N10N11energyscan} for a full scan of the variational energy at $N_e=10$ and $N_e=11$ on a square lattice, and Fig.~\ref{fig:N10energyscanHexagon} for $N_e=10$ on a hexagonal and $\tau=2\i$ rectangular lattice.

To generate the CF energy landscape in the figure above we have performed a search over FS configurations and found the ones that minimizes (\ref{eq:cf_energy}) in each momentum sector.
See App.~\ref{sec:Algorithm} for an algorithm that generates all FS configurations in order of monotonically increasing CF energy.
There will typically be more than one FS configuration that minimizes
(\ref{eq:cf_energy}). Some of these will be related by symmetry transformations such as reflections or rotations, but others may actually represent different shaped Fermi discs. These accidental degeneracies can usually be split by the surface tension, mentioned earlier.

  Within a momentum sector, the ground state configuration will be given by the most closely packed CF state that still respects the momentum constraints. As the effective origin depends on the momentum sectors, the momentum sector of the global ground state will depend strongly on the system size as well as shape of the torus. See Tab.~\ref{tab:GS-states} for a list of momentum sectors containing the global grounds states at various system sizes.

\section{particle-hole and Orbital Inversion Symmetries on the torus}\label{sec:PH_inv}
The LLL is exactly particle-hole symmetric as long as LL mixing is ignored but the CF construction
is not PH-symmetric.
The consensus is however that CFs restore PH symmetry approximately, something that is now also observed for bosons\cite{Geraedts_2017}. 
It is interesting to ask to what extent this approximation holds for the FS configurations.
In this setting it is also interesting to consider to which degree orbital inversion symmetry is respected,
as inversion is also a symmetry of the LLL (one that does not change the filling fraction).

We now consider the behavior of the trial  wave functions under two types of transformations on the torus: particle hole symmetry $C$ and orbital inversion symmetry $I$. These are defined as follows,
\[
C=\mathcal{K}\prod_{n=1}^{N_{s}}\left(a_{n}^{\dagger}+a_{n}\right)
\]
where $\mathcal{K}$ is complex conjugation and $a^\dagger_n$ and $a_n$ are the electron creation and annihilation operators for orbital $n$.

The other transformation is an orbital reflection (or inversion) that exchanges the $i^{\rm th}$ and $(N_{s}-i)^{\rm th}$ orbital in Landau gauge.
In terms of operators, inversion  is defined as
\[
I=\prod_{i=1}^{\left[N_{s}/2-1\right]}I_{i,N_{s}-i+1},
\]
where the operator acting on the sites $i$ and $j$ can be written as 
\[
I_{ij}=a_{i}^{\dagger}a_{j}+a_{i}a_{j}^{\dagger}+1-N_{i}-N_{j}+2N_{i}N_{j}.
\]

The operator $C$ has commutation relation
\begin{equation}
  CT_j=(-1)^{N_s-1}T_jC\label{eq:CT}
\end{equation}
for $j=1,2$ and also changes the filling fraction as $\nu\to1-\nu$.
Inversion acts non-trivially on $T_1$ and on $T_2$ as
\begin{equation}
  IT_2=T_2^{-1}I\label{eq:IT}
\end{equation}
but does not change the filling fraction $\nu$.
Note that particle-hole conjugation and inversion commute ($IC=CI$).

For a well defined momentum $K_{1}$ at a general filling fraction, the two operators have the effect
\begin{eqnarray*}
K_{1} & \overset{C}{\to} & \sum_{j=1}^{N_{s}}j-\sum_{j=1}^{N_{e}}k_{j}=\frac{N_{s}\left(N_{s}+1\right)}{2}-K_{1}\\
K_{1} & \overset{I}{\to} & \sum_{j=1}^{N_{e}}\left(N_s+1-k_{j}\right)=N_{e}-K_{1}
\end{eqnarray*}

Let us now specialize to $N_{s}=2N_{e}$, giving $K_1 \overset{C\,\mathrm{or}\,I}{\to} N_e-K_1$.
To appreciate the effect on $K_2$ by $I$ and $C$, it is instructive to consider the effect of $I$ and $C$ on a state with a well defined $K_{2}$ quantum number.
The eigenvalues $K_2$ are then obtained as
\begin{eqnarray*}
T_2^2 (I \ket{K_2}) &=& I e^{-\i2\pi\frac{K_2}{Ne}} \ket{K_2} = e^{-\i2\pi\frac{K_2}{Ne}} (I \ket{K_2})\\
T_2^2 (C \ket{K_2}) &=& C e^{\i2\pi\frac{K_2}{Ne}} \ket{K_2} = e^{-\i2\pi\frac{K_2}{Ne}} (C \ket{K_2})
\end{eqnarray*}
where we use the commutations relations \eqref{eq:CT} and \eqref{eq:IT}.
In both cases $K_2 \overset{C\,\mathrm{or}\,I}{\to} -K_2$. We note here that for a half filled system $C$ and $I$ map between the same two sectors, but they do it in different ways and in general $I\neq C$.

We also note that in general, momentum $\mathbf K$ is not preserved under $C$ and $I$ which means that it is often wrong to talk about particle-hole/inversion ``invariance''.
Rather one should be talking about ``co-variance'' under these transformations.
There is an important exception to this picture, the self-dual sectors where $K_{j}=-K_{j}$ or $K_{j}=N_{e}-K_{j}$.
These sectors are $\left(K_{1},K_{2}\right)=\left(0,0\right)=\left(0,\frac{N_{e}}{2}\right)=\left(\frac{N_{e}}{2},0\right)=\left(\frac{N_{e}}{2},\frac{N_{e}}{2}\right)$. Since $C$ and $I$ act within these sectors,  we can really check for invariance here.

A first question to investigate is how invariant/covariant the FS states are, and how different FS states map to each other under PH/inversion.
To answer this question we fix a small system size (for illustrative purposes) $N_{e}=10$  and consider the $N_{FS}=30$ FS configurations with the lowest total CF-energy in a few selected symmetry sectors for hexagonal and square geometry.
See Fig.~\ref{fig:PH_vs_Inversion_invariance}.
In the self dual sector $\left(5,5\right)$ we then plot the overlap between all of the $N_{FS}$ states and their PH/inverted counterparts.
In the non-self dual sectors we must construct one set of $N_{FS}$ states in sector $\mathbf{K}$ and then another set of $N_{FS}$ states in sector $-\mathbf{K}$ that also minimize the CF energy.
The PH/inversion then transforms the sector $-\mathbf{K}$ into the sector $\mathbf{K}$ which enables us to take overlaps between all states in the usual manner. 

In the square geometry, we find consistently good overlap of the PH/Inversion images of these low CF energy states and the reversed flux duals of the same states.
At zero effective flux, the reverse flux attachment construction is equivalent to the direct flux attachment construction and means that, when taking the reversed flux dual of a state, the FS set $\left\{\mathbf k^j\right\} $ is replaced
by $\left\{ -\mathbf k^j\right\} $ (\ie we rotate it by $180{}^{\circ}$).
The need to rotate the FS configuration $180{}^{\circ}$ persists even in the self dual sectors.
We also find a general trend that the FS states are more symmetric under inversion than under PH-symmetry but the effect is small.

We also performed the same test on hexagonal (See again Fig.~\ref{fig:PH_vs_Inversion_invariance}) as well as rectangular tori. We find the same general picture, namely that FS configuration tend to have high overlap with their reversed flux counterparts. For some states the actual overlaps are not particularly large here, but the overlap between any FS state and its reversed flux dual still tends to be larger than the overlaps with the other CF states we considered.

\section{Quality of projected trial wave functions}\label{sec:Proj_wfn}
\begin{figure*}[htb]
  \begin{center}
    \includegraphics[width=1\linewidth]{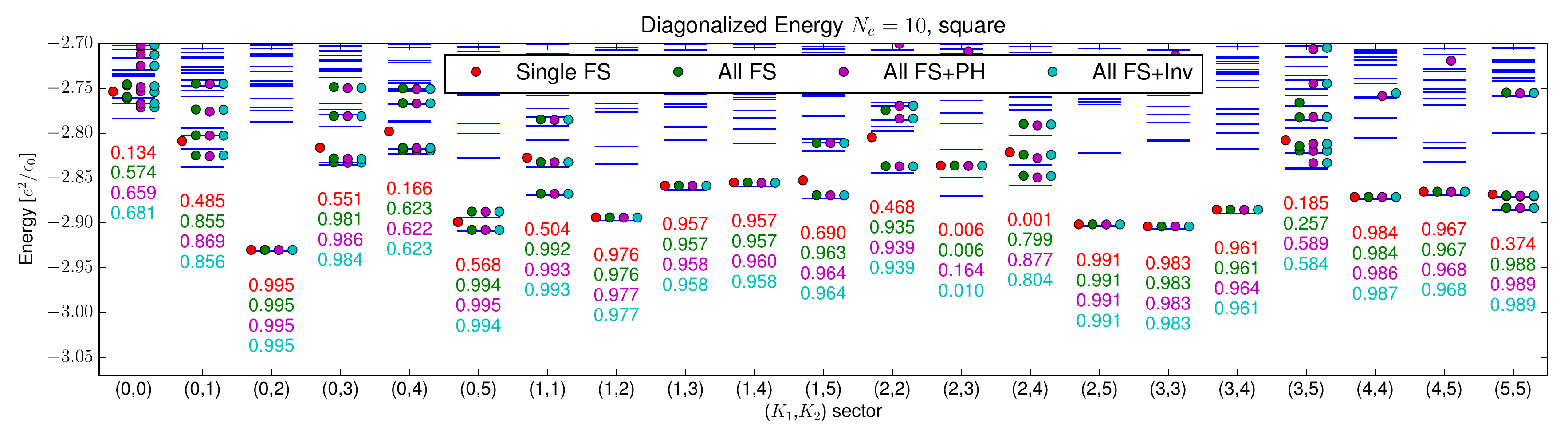} \\  
  \includegraphics[width=1\linewidth]{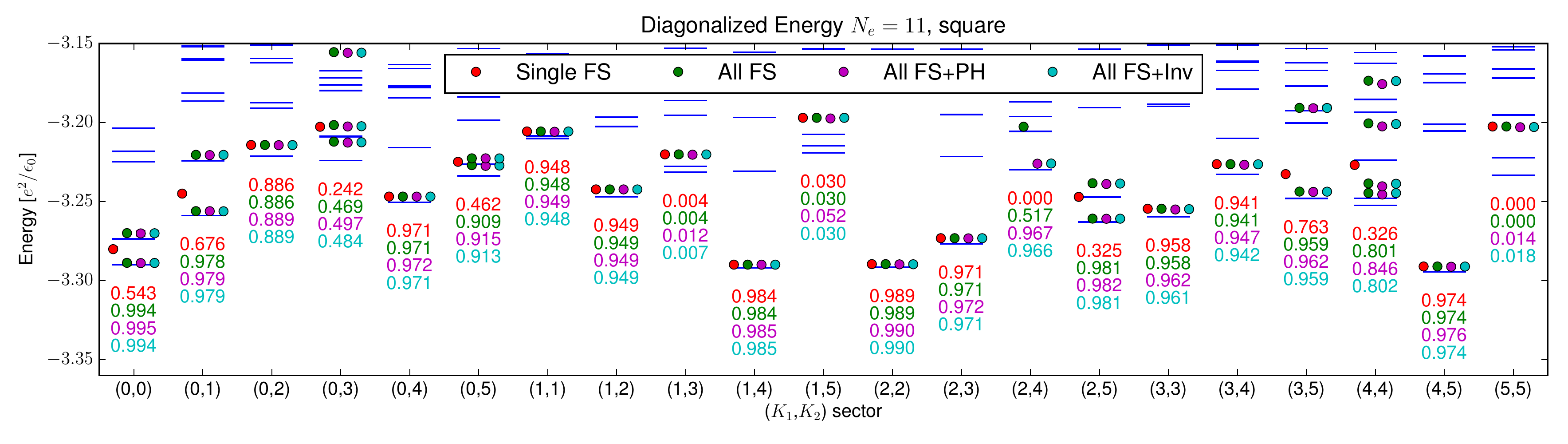} 
  \end{center}
  \caption{Comparing variational energy of FS states to the exactly diagonalized energy for $N_e=10$ (upper) and $N_e=11$ (lower) on a square torus in all momentum sectors (up to $C_4$ rotations).
    The blue horizontal bars show the exact energy of the Coulomb interaction from diagonalization.
    The red discs are the variational energy of (one of) the FS configuration(s) that minimize the CF energy within a given $(K_1,K_2)$ sector.
    The green discs are the eigen-energies of the Coulomb interaction, projected onto the space spanned by all FS configurations (related by symmetry) that minimize the CF energy.
    The purple/cyan discs are eigen-energies of the Coulomb interaction, projected onto the space spanned by all FS configurations (related by symmetry) that minimize the CF energy as well as their PH conjugated/orbital inverted reversed flux duals.
    The colored numbers in each momentum sector show the squared overlap between the lowest energy FS superposition (of the corresponding color) and the Coulomb ground state in that momentum sector.
    As can be seen, the variational energy of the chosen FS configurations (green) is close to the Coulomb energy, and the squared overlap is almost always very close to one.
    The larger overlaps are found in momentum sectors with lower energies and where the energy gap is large.
    Adding in PH-conjugation (purple) or orbital inversion (cyan) has most often negligible impact on the variational energy and overlap with the Coulomb ground state.
    This is related to the fact that most FS configurations are highly PH/inversion symmetric, see Fig.~\ref{fig:PH_vs_Inversion_invariance}.}
  \label{fig:N10N11energyscan}
\end{figure*}
\begin{figure*}[htb]
  \begin{center}
    \includegraphics[width=.539\linewidth]{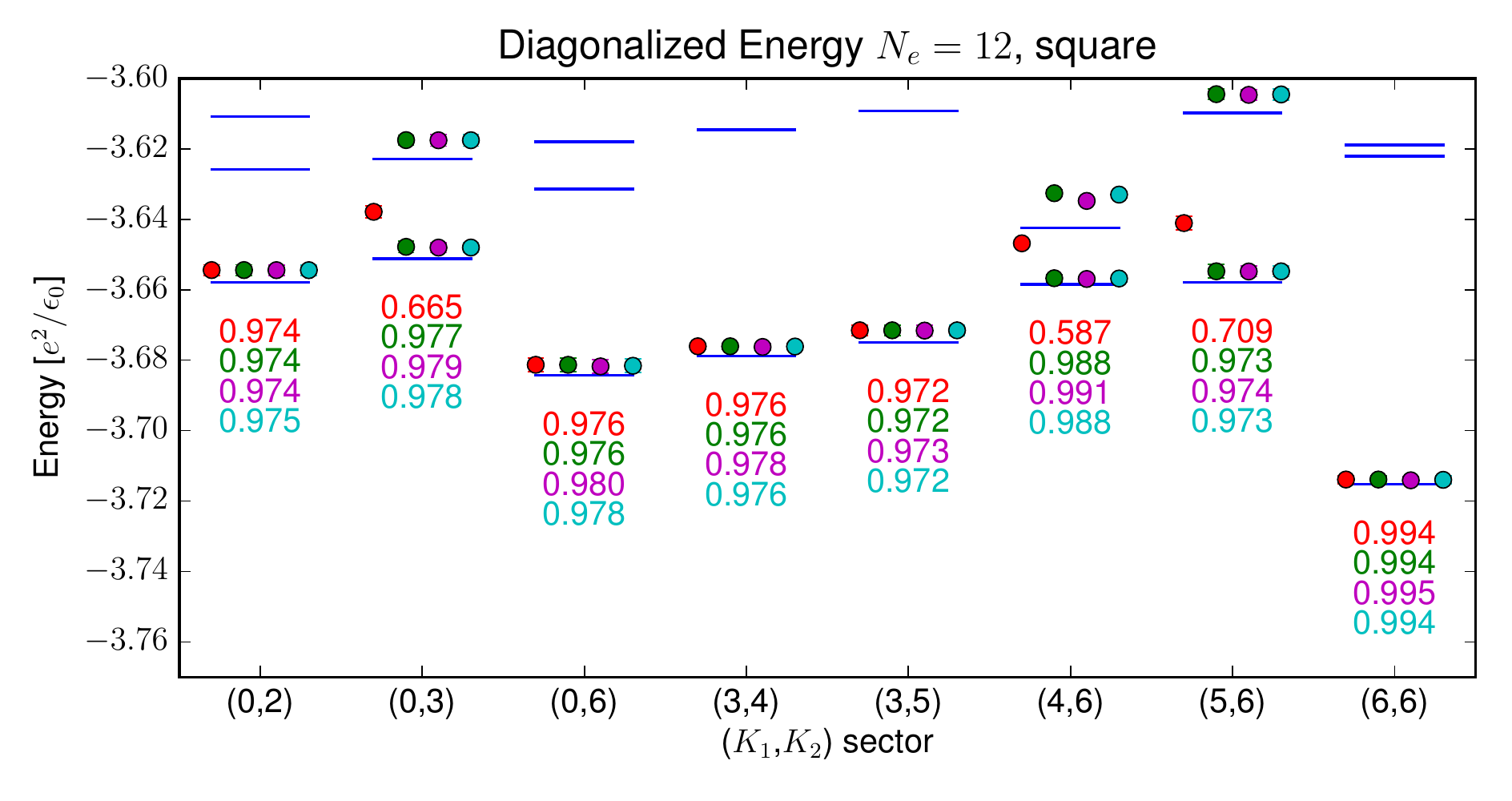}
  \includegraphics[width=.271\linewidth]{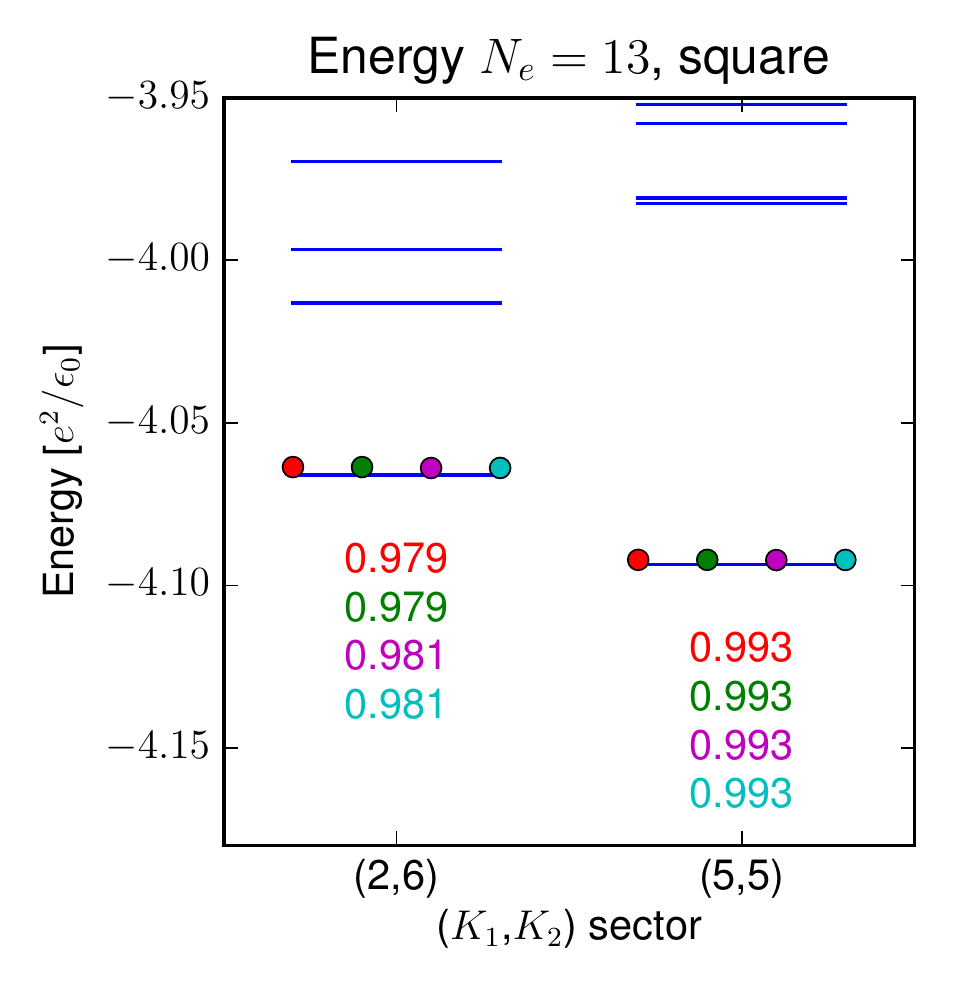} 
  \end{center}
  
  \caption{
    Comparing variational energy of FS states to the exactly diagonalized energy for $N_e=12$ (left) and $N_e=13$ (right) on a square torus for the lowest energy momentum sectors (up to $C_4$ rotations).
    The symbols are as in Fig.~\ref{fig:N10N11energyscan}.
    Just as in Fig.~\ref{fig:N10N11energyscan}, if there are more than one FS configurations that minimize the CF energy all of the FS states need to be taken into account to obtain a good approximation of the Coulomb ground states.
    Also here, PH conjugation/inversion has a minute effect on the wave function.
\label{fig:N12N13energyscan}}
\end{figure*}
\begin{figure*}[htb]
  \begin{center}
    \includegraphics[width=1\linewidth]{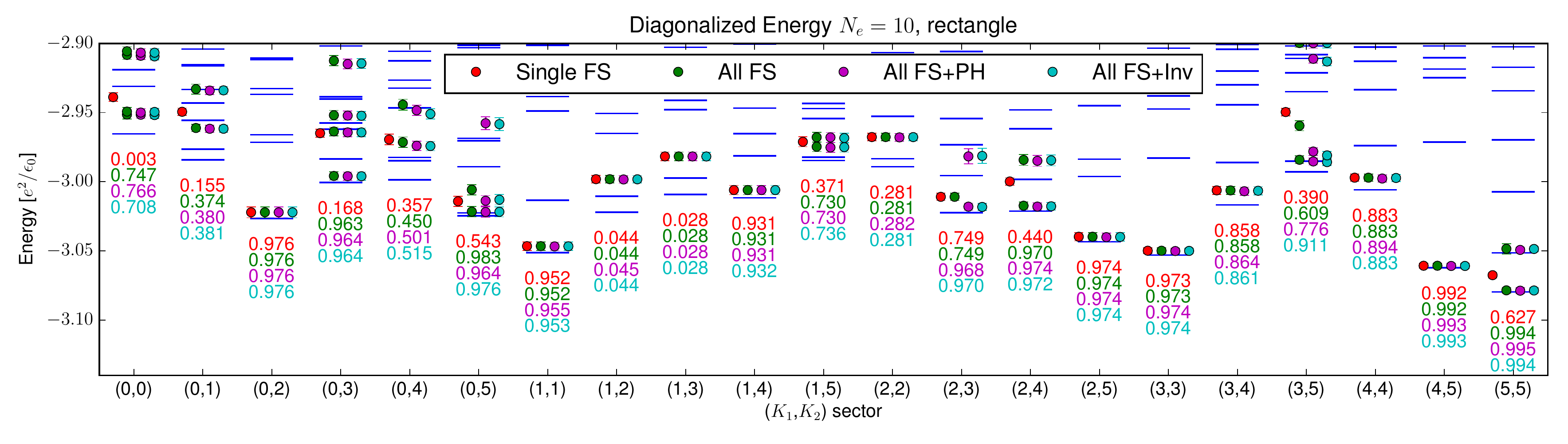} \\  
  \includegraphics[width=1\linewidth]{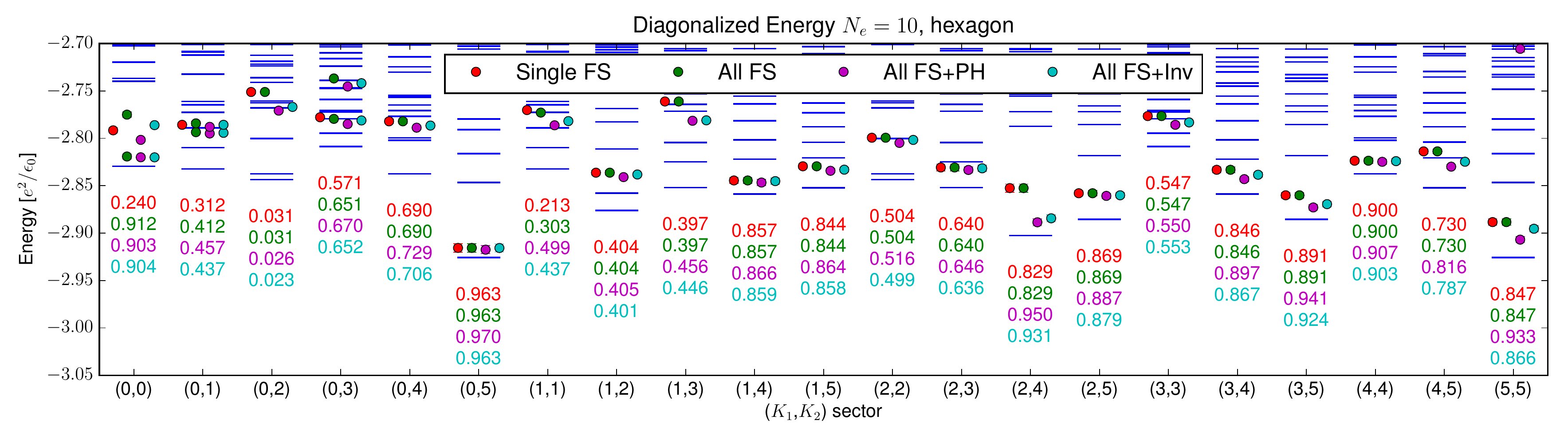} 
  \end{center}
  \caption{
    Comparing variational energy of FS states to the exactly diagonalized energy for $N_e=10$  on the $\tau=2\i$ rectangular torus (upper) and the hexagonal torus (lower).
    The symbols are as in Fig.~\ref{fig:N10N11energyscan}.
    Here we see that the composite fermion wave function is generally less good at approximating the Coulomb ground state, than was the case on the square torus.
    On the rectangular torus, adding in PH/inversion symmetry does little to improve the wave functions, but on the hexagonal tours a definite improvement can bee seen in some momentum sectors, \eg $(2,4)$ and $(5,5)$.
    This is not surprising given that the FS configurations where less PH/inversion symmetric on the hexagonal torus than on the square torus.
    See Fig.~\ref{fig:PH_vs_Inversion_invariance}.}
  \label{fig:N10energyscanHexagon}
\end{figure*}

We now wish to test the quality of the CF wave functions (\ref{eq:CF_wf}) against the exact ground state in each momentum sector.
First we find the trial states with the lowest CF energy (there may be multiple states related by symmetry).  We project these trial states onto the LLL using energy projection\cite{Fremling_16} and orthogonalize the resulting states. All diagonalization is performed using the Hammer package\footnote{http://www.thphys.nuim.ie/hammer/}. Finally, we diagonalize the Coulomb interaction within the trial space. This yields the variational energies shown in figures~\ref{fig:N10N11energyscan},~\ref{fig:N12N13energyscan} and \ref{fig:N10energyscanHexagon}

We will report here on the overlap between FS states for all momentum sectors on a square torus for $N_e=10$ and $N_e=11$ (see Fig.~\ref{fig:N10N11energyscan}) and will show selected sectors for $N_e=12$ and $N_e=13$ (see Fig.~\ref{fig:N12N13energyscan} ).
We also report on $N_e=10$ for a hexagonal and rectangular torus (see Fig.~\ref{fig:N10energyscanHexagon}).
It is in principle possible to push the numerics to larger system sizes, but at an exponentially increasing cost.

\subsection{Individual FS states}
We begin by considering how well the restriction of minimizing CF energy and surface tension works when considering the variational energy of individual FS states.
In Fig.~(\ref{fig:N10N11energyscan}) we see the variational energy of (one of) the pre-selected FS-states shown in red.
We see that the variational energy of this state displays features that are similar to the exact energy landscape.
We note that for the lowest exact energy states the variational energy
is almost the same as the exact energy with overlaps that are
close to unity at $\left(K_{1},K_{2}\right)=\left(0,2\right),\left(1,2\right),\left(2,5\right),\left(3,3\right)$
with squared overlap of $0.995$, $0.976$, $0.991$, $0.983$ respectively for $N_{e}=10$.
At $N_{e}=11$ the story is similar with the lowest energy sectors at $\left(K_{1},K_{2}\right)=\left(1,4\right),\left(2,2\right),\left(4,5\right)$
showing squared overlaps of $0.984$, $0.989$, $0.974$ respectively. 

There are some notable exceptions to the above mentioned story such as $\left(0,5\right)$ and $\left(5,5\right)$ at $N_{e}=10$, $\left(0,0\right)$ and $\left(2,5\right)$ at $N_{e}=11$.
The squared overlap here is only $0.568$, $0.374$, $0.543$ and $0.325$ respectively.
In these cases there also exist at least one other symmetry related state that must be taken into account.
Adding these symmetry partners and diagonalizing in the space spanned by these extra states (green)
we obtain ground state overlaps that are $0.995$, $0.988$, $0.994$ and $0.981$ with the lowest energy state.
For the larger systems at $N_{e}=12$ and $N_{e}=13$ the story is the same:
If the pre-selected FS state is unique it has a large overlap with the ground state,
and if it isn't unique then adding the symmetric cousins yields squared overlaps that are above $0.97$.

Further, we observe that states with high energy are typically less well described by the the CF states. This also correlated with the energy gap being smaller when the CF-description is worse, implying the need to include CF interactions at higher energy.

\subsection{PH/Inversion symmetrization}
Next we consider the effect of including particle-hole and inversion transformations into the trial space.
We are interested in investigating if making the wave functions PH symmetric has an effect on the overlap with the Coulomb ground state.
We proceed as follows: If the FS states at momentum $\mathbf{K}$ are given by the configurations \{$\mathbf{k}$\}, we simply construct the reversed flux duals of these CF states with configurations \{$-\mathbf{k}$\} in the $-\mathbf{K}$ sector.
We then PH-conjugate/invert these CF states to the momentum sector $\mathbf{K}$ and add them to our trial space, and diagonalize the Coulomb interaction within this space.

The data for PH-conjugation and inversion can be found in the purple and cyan data respectively in Fig.~\ref{fig:N10N11energyscan} and \ref{fig:N12N13energyscan}.
In the figures we see that including PH/inversion doubles the number of states in the trial space (as it should) but it has almost negligible effect on both the variational energy and the overlap with the ground state.
We note that when there is a difference between PH and inversion,
it is as a rule PH conjugation that gives the best (but still small) boost in overlap.
This fits well with the observation in Section \ref{sec:PH_inv} that the FS states tended to be more inversion covariant than PH covariant.
The reader may note that the actual numbers of visible states in purple/cyan is not always twice the number of visible green states.
These states can be found at higher energy than visible and are dominated by MC noise, as the FS states are almost completely PH/inversion covariant.

The energy projection method provides very precise results for the overlaps and variational energies of states with adequate weight in the LLL and reasonable overlap with the low lying Coulomb spectrum, provided that sufficiently many MC samples are used.
See Ref.~\onlinecite{Fremling_16}, for details.
Here these conditions are satisfied and the errors on the variational energy  (included in the figures) are smaller than the colored discs in all cases except potentially for some of the highest energy states considered.
The dominant (small) systematic error in the method actually causes the overlaps of good variational states to be underestimated and their variational energies to be overestimated.
This effect is due to an overestimation of overlaps on the high energy, low overlap states, which causes overlaps on good trial states to be renormalized downwards (and variational energies to be pushed upwards). 

\subsection{Hexagon and Rectangle}
We also considered the hexagonal torus as well as rectangular torus at $\tau_{2}=2$, see Fig.~\ref{fig:N10energyscanHexagon} for data.
We find that in both geometries the FS-states are a worse match to the exact ground state than in the square geometry.
However for the very lowest energy states
-- $\left(0,5\right)$, $\left(2,4\right)$, $\left(2,5\right)$ for hexagonal,
$\left(5,5\right)$, $\left(4,5\right)$, $\left(3,3\right)$ , $\left(2,5\right)$ $\left(1,1\right)$ for rectangular --
the FS - description still works well.
The FS-states are better for the rectangular than the hexagonal lattice.

We note that also here, just as in the square case, neither PH conjugation nor inversion symmetry has a large impact on the improvement of the trial space.
The FS states are already close to PH/inversion covariant.
Again, the improvement is slightly better for PH than inversion.

\begin{figure}
  \begin{center}
    \includegraphics[width=1\linewidth]{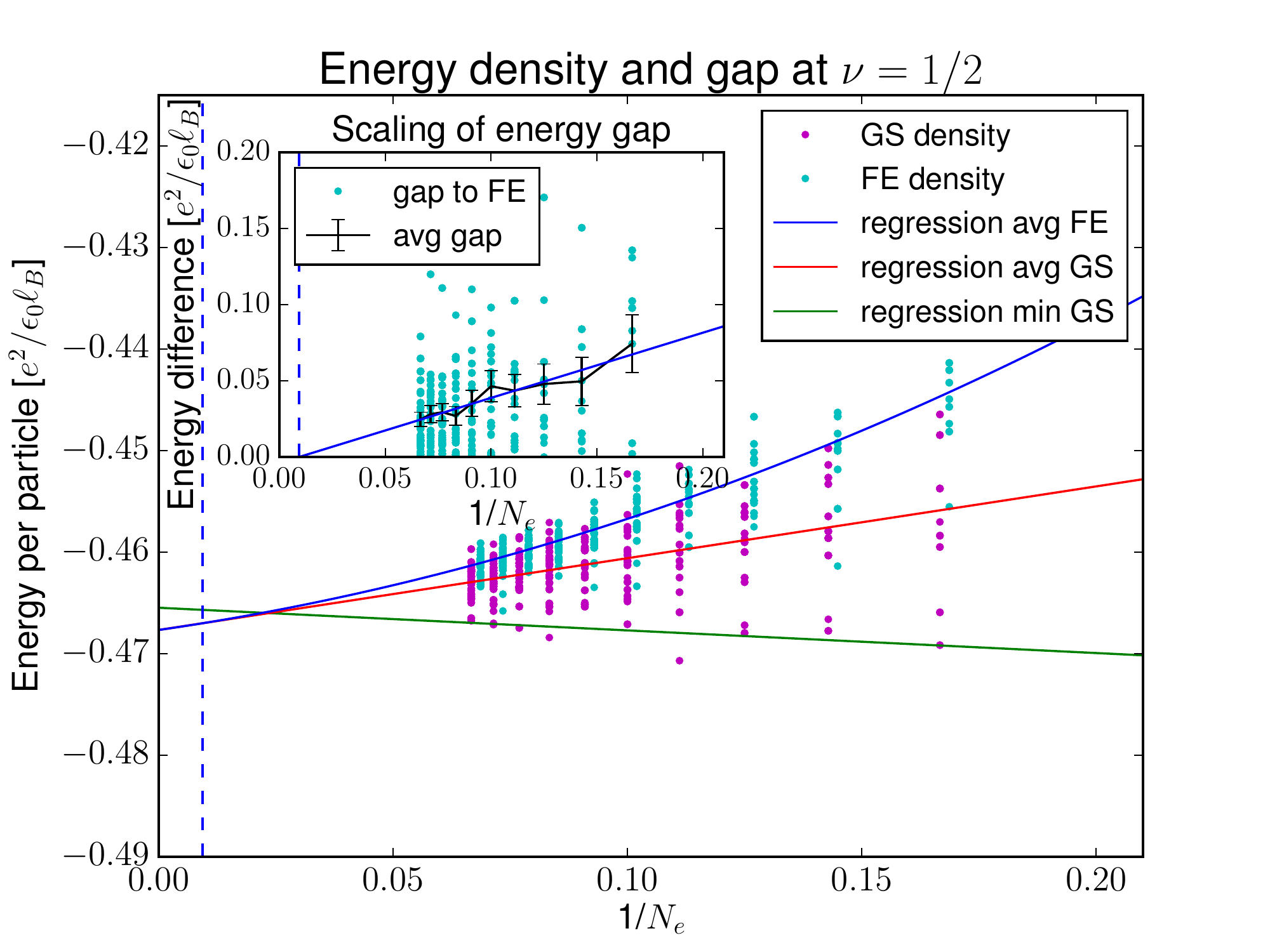}
  \end{center}
  \caption{Extrapolation of the exact average ground state (GS) energy and average (momentum preserving) first excited (FE) state for a square torus to the thermodynamic limit. Data for hexagon looks qualitatively the same.\\
    {\bf Inset:}  The gap (cyan) in all momentum sectors, and the gap averaged over all momentum sectors  (black bars) extrapolated  to the thermodynamic limit (blue line).
    We conclude that the average gap is is closing in the thermodynamic limit.\\
    {\bf Main Plot:} The ground state (purple) and first exited state (cyan) energy density for all momentum sectors.
    Average (red) and global (green) ground state energy extrapolated to the thermodynamic limit.
    The blue line is the gap size extracted from the inset superimposed on the average ground state energy (red).
    Note: The horizontal position of the energies of the GS and FE states are slightly shifted with respect to each other for better readability.}
  \label{fig:scaling_gap}
\end{figure}

\subsection{Thermodynamic limit}
Finally we verify that the expected CF Fermi liquid state at $\nu=\frac12$ really is gapless, by extrapolating the ground state energy density to the thermodynamic limit.
As the structure of the energy landscape is non-trivial -- with degeneracies and a constantly moving global ground state -- we choose to focus on the gap for momentum preserving excitations, averaged over all momentum sectors.
We find (inset of Fig.~\ref {fig:scaling_gap}) that the gaps do scale to zero in the thermodynamic limit.
We also extrapolate the energy density of the global ground state (green) and find a thermodynamic value of $\epsilon=-0.4655\frac{e^2}{\epsilon_0\ell_B}$.
If we instead extrapolate the average energy of the lowest energy states (red) over all momentum sectors we obtain $\epsilon=-0.4677\frac{e^2}{\epsilon_0\ell_B}$.
Both of these values agree reasonably well with the energy density obtained by scaling the exact energy of the $N_s=2 N_e\pm1$ states (studied in more detail section \ref{sec:Charged_excitations}). For those states, the results are $\epsilon_+=-0.4650\frac{e^2}{\epsilon_0\ell_B}$
and $\epsilon_+=-0.4639\frac{e^2}{\epsilon_0\ell_B}$, see Fig.~\ref{fig:2Ne_pm_1_var}.

\section{Charged excitations}
\label{sec:Charged_excitations}
In this section we consider adding charged excitations on top of the half filled state in the sense of adding or reducing one flux quantum for at total of $N_{s}^{\pm}=2N_{e}\pm1$ fluxes, giving a filling fraction $\nu_{\pm}=\frac{N_e}{2N_{e}\pm1}$.
In this setting the composite fermions will be moving in an effective magnetic field reduced to a single flux quantum.
The composite fermions will then occupy a tower of $\Lambda$-levels with only a single electron in each level.
The CF wave function is therefore written as 
\begin{equation}
\psi_{\nu_\pm}=P_{\mathrm{LLL}}\det\left(\prod_{j=0}^{N_{e}-1}\eta_{n_{j}}^{\left(\pm\right)}\left(z_{j}\right)\right)\phi_{\frac{1}{2}}\label{eq:CF_2Ne_pm_1}
\end{equation}
where $\eta_{n}^{\left(+\right)}\left(z\right)$ is the (only) $n^{\rm th}$
level Landau orbital at $N_{\phi}=1$ and
$\eta_{n}^{\left(-\right)}\left(z\right)=\left(\eta_{n}^{\left(+\right)}\left(z\right)\right)^{\star}$
is the complex conjugate for reverse flux attachment.

At this filling fraction, $T_1$ only commutes with $T_2^{2N_e\pm1}$ which
means that there exists only one $K_2$ sector and it has quantum number $K_2=0$.
All the $K_1$ momentum sectors are degenerate and related by $T_2$ translation operations.
The wave function $\psi_{\nu_\pm}$ in (\ref{eq:CF_2Ne_pm_1}) will not be in a well defined momentum state,
but we may without loss of generality project it onto the $K_{1}=0$ momentum sector as 
\[
\psi_{\nu_\pm,K_{1}=0}=\frac{1}{\sqrt{N_{s}^{\pm}}}\sum_{j=1}^{N_{s}^{\pm}}T_{1}^{j}\psi_{\nu_\pm}.
\]

\begin{figure}
\begin{centering}
\includegraphics[width=1\linewidth]{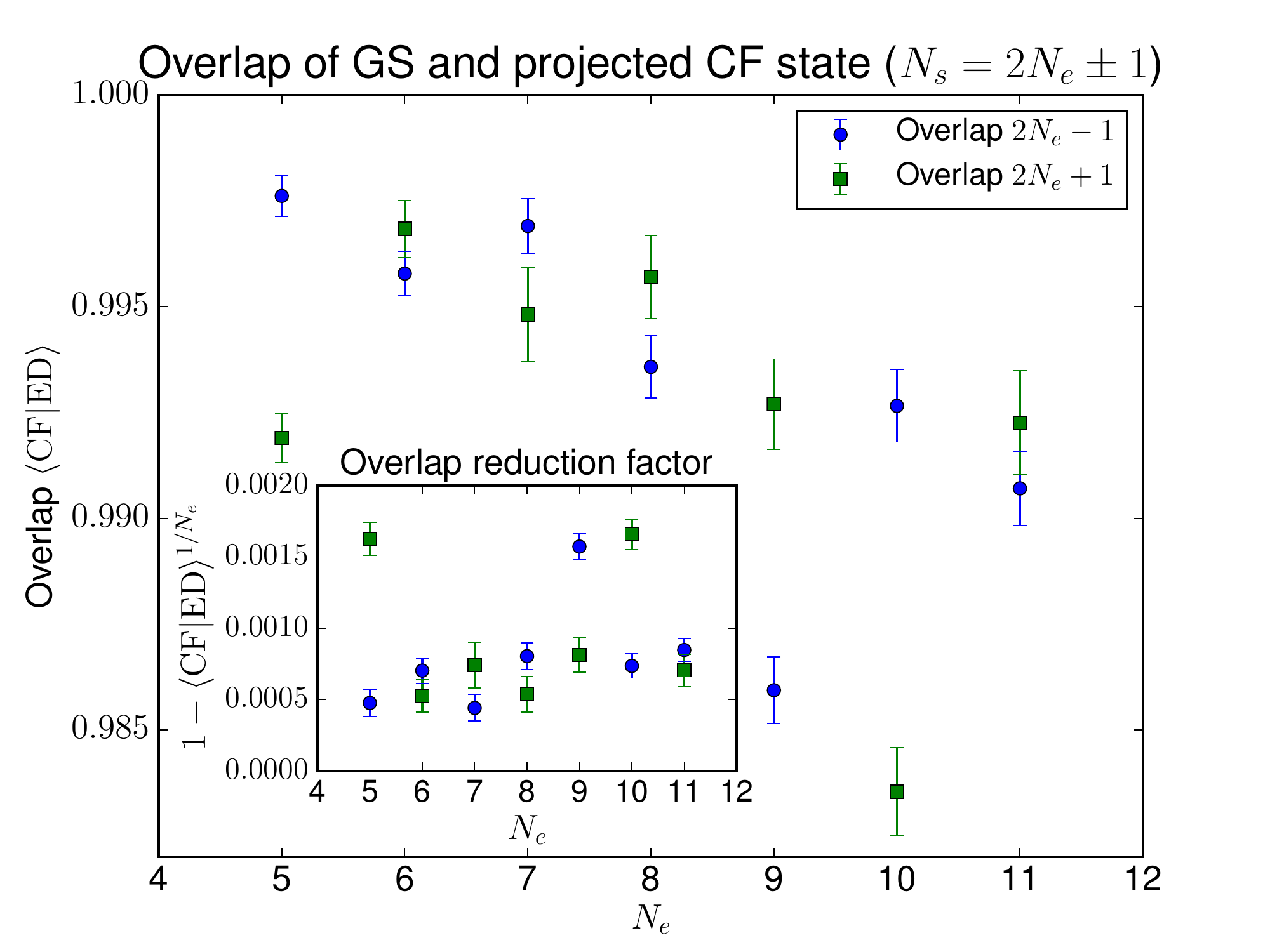}
\par\end{centering}

\caption{
  {\bf Main Plot:} The overlap between the Coulomb ground state at $N_{e}$ electrons
and $N_{s}=2N_{e}\pm1$ flux quantum and the CF wave function (\ref{eq:CF_2Ne_pm_1})
for that flux.
The overlap is higher than 0.99 for all system sizes considered, except for two outliers at $N_e=9,10$.
here seems to be an even odd effect as to whether CF($\pm$) has higher overlap at a given system size.
\\
  {\bf Inset:} Overlap reduction factor per particle $\epsilon$, related to the overlap as $\left|\braket{\psi_{CF}}{\psi_{ED}}\right|=\left(1-\epsilon\right)^{N_e}$. This is stable at $\epsilon\sim8\cdot10^{-3}$ almost independently of system size.}
\label{fig:2Ne_pm_1_ov}
\end{figure}

\begin{figure}
\begin{centering}
\includegraphics[width=1\linewidth]{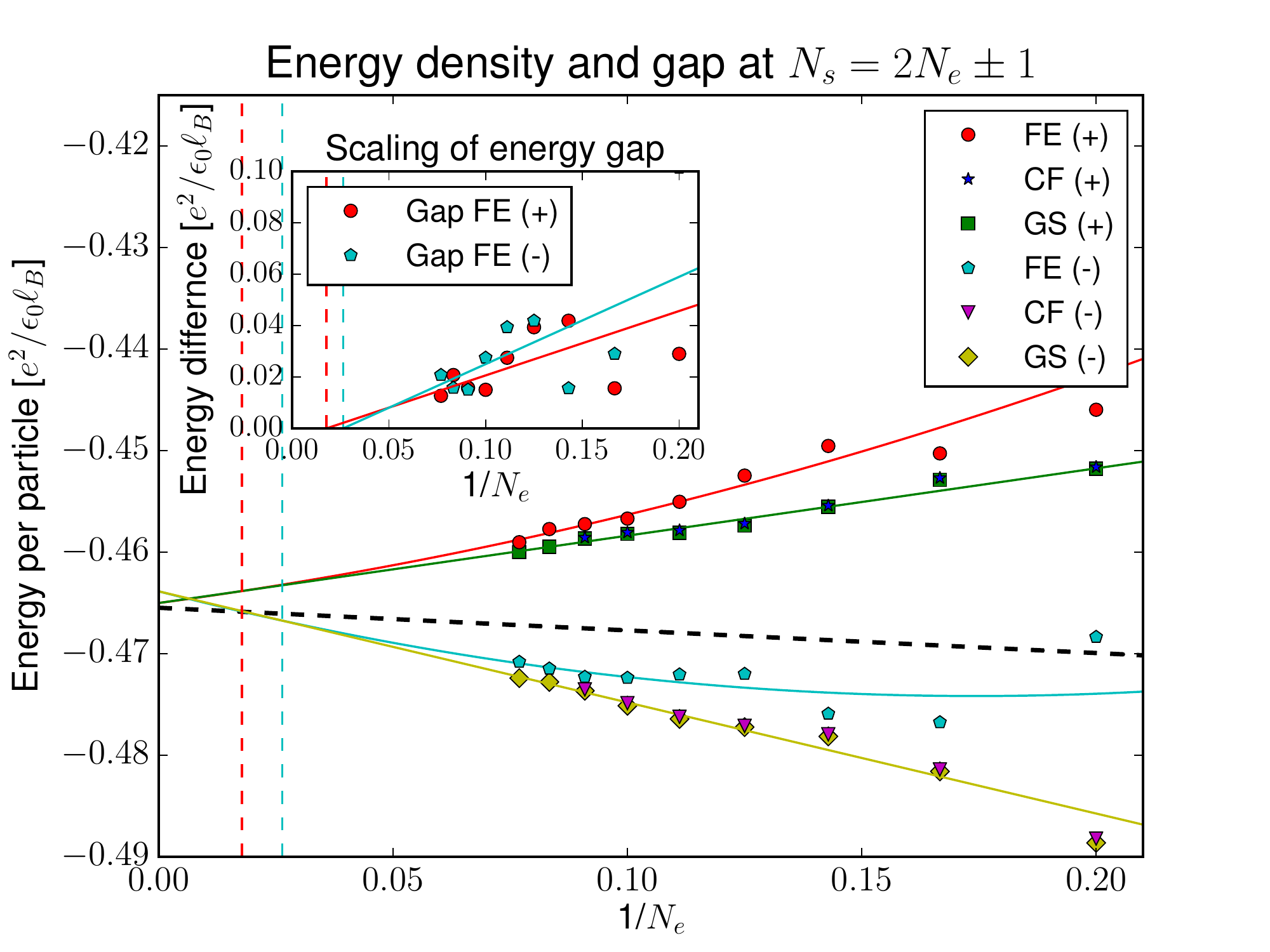}
\par\end{centering}
  \caption{
    {\bf Inset:}  The exact gap at $N_s=2N_e+1$ (red) and $N_s=2N_e-1$ (cyan) extrapolated  to the thermodynamic limit for a square torus.
    The gap appears to be closing in the thermodynamic limit, both from higher and lower number of fluxes, but we are not yet in a scaling regime.\\
    {\bf Main Plot:} The energy density of the exact ground state (GS), first excited (FE) state and composite fermion state (CF) for a square torus with $N_s=2N_e\pm 1$, and their extrapolations to the thermodynamic limit.
    The red and cyan lines are the gap size extracted from the inset, superimposed on the ground state energies (green and yellow).
    The composite fermions model the ground states very well and variational energy of the composite fermion states are almost indistinguishable from the energies of the exact ground states on the scale of the plot.
    The black dashed line is the extrapolation of the global ground state energy for a half filled system (same as green line in Fig.~\ref{fig:scaling_gap}).
  }
  \label{fig:2Ne_pm_1_var}
\end{figure}

We again use energy projection to approximate the $P_{\mathrm{LLL}}$ operation.
We first note that the CF state describes the Coulomb ground state very well,
with an overlap that is above 0.99 for all system sizes considered, see Fig.~\ref{fig:2Ne_pm_1_ov}.
The overlap is naturally falling, so in the inset we also plot the effective
overlap reduction factor per particle $\epsilon=1-\left|\braket{\psi_{CF}}{\psi_{ED}}\right|^{\frac{1}{N_{e}}}$,
which is stable at $\epsilon\sim8\cdot10^{-3}$ almost independent of system size.

We also consider the variational energy of the CF state compared to the exact energy of the ground state as well as the first excited state, see Fig.~\ref{fig:2Ne_pm_1_var}.
We find that the variational energy approached from both higher and lower fluxes converge to the same value,
which is consistent with the value obtained for the scaled ground state energy precisely at half filling -- see Fig.~\ref{fig:scaling_gap}.
We also find that the gap between the first excited and ground state decreases as the system size increases.
Here it does however seems that we are not yet in a scaling regime,
even though it looks plausible that the gap vanishes in the thermodynamic limit.
One should remember here, when comparing to the half filled result,
that due to the smaller symmetry, the Hilbert space that needs to be constructed for diagonalization is roughly $N_e\cdot2^{\pm1}$ times the size of the Hilbert space at half filling.
As a direct consequence we cannot push the numerics to the same system sizes.

\subsection{Charge gap in the thermodynamic limit}
To determine the charge gap in the thermodynamic limit, we wish to estimate the energy needed to add or remove a magnetic flux, while preserving the area of the torus.
From the scaling relations in Fig.~\ref{fig:scaling_gap} and Fig.~\ref{fig:2Ne_pm_1_var}
it should be clear that the energy per particle of a system with $N_s=2N_e+s$ fluxes,
$s=-1,0,1$ is approximately given by $E_s/N_e=c+d_s/N_e$, where $c$ is independent of $s$.
The total energy  can thus be written as 
\[E_{s} = (\varepsilon_s -\varepsilon \cdot N_e)\frac{e^2}{\epsilon_0\ell_B},\]
where $\varepsilon_s$ is the slope, and $\varepsilon$ is the (negative) intercept in the figures.
When computing the charge gap, we must take care to ensure that the torus area is kept constant, which amount to letting $\ell_B$ depend on $N_s$ through $A=2\pi N_s \ell^2_B$.
Taking this into account gives the charge gap
\[ \Delta_\pm = E_{\pm} - E_{0} \approx \left(\varepsilon_\pm - \varepsilon_0 \mp \frac{\varepsilon}4\right)\frac{e^2}{\epsilon_0\ell_B},
\]
as compared to $(\varepsilon_\pm-\varepsilon_0)\frac{e^2}{\epsilon_0\ell_B}$ if $\ell_B$ is held constant.

\begin{figure}
\begin{centering}
\includegraphics[width=1\linewidth]{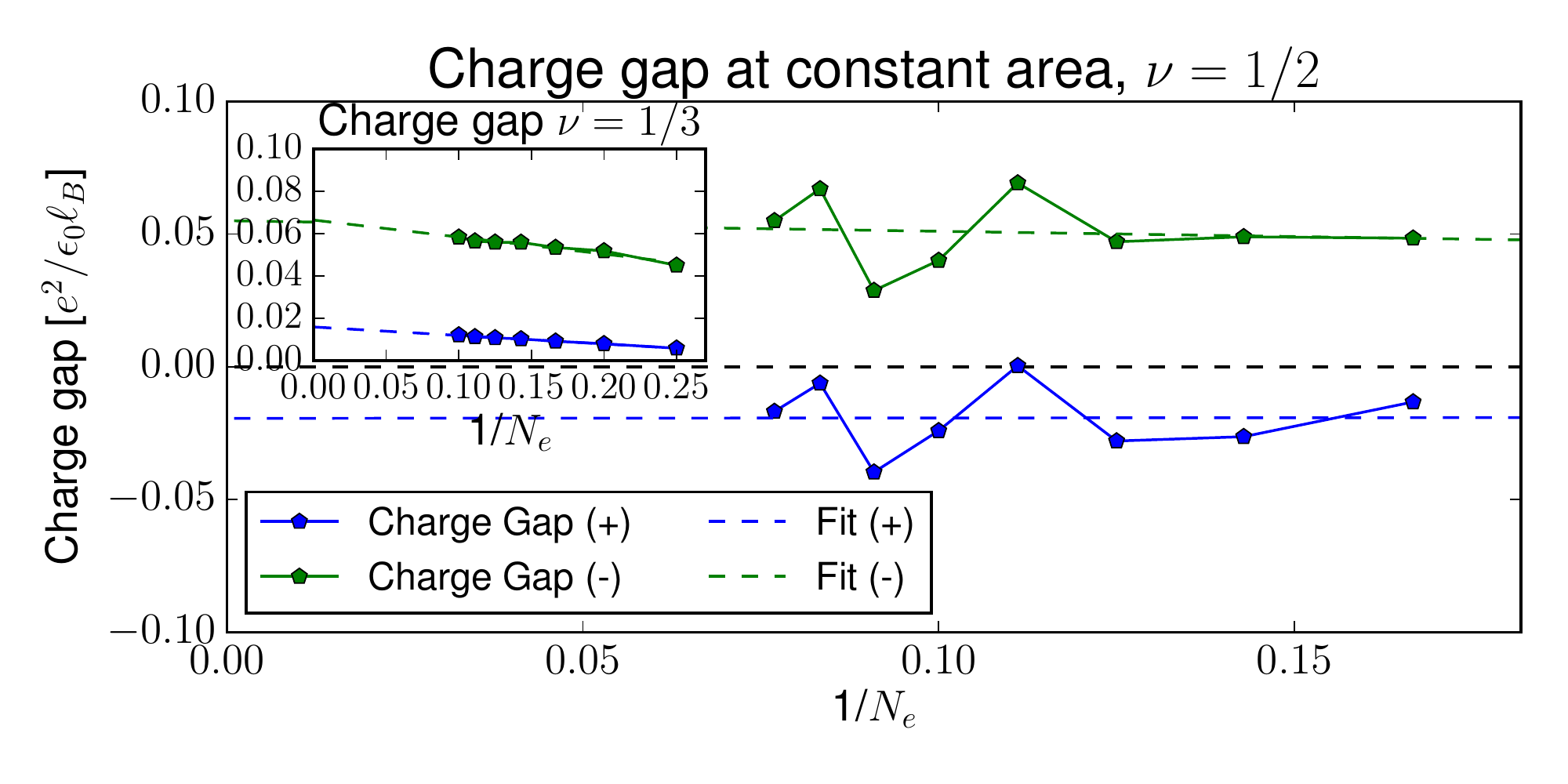}
\par\end{centering}
\caption{
  {\bf Main Plot:}
  The charge gap at constant torus area $A$ (as opposed to constant magnetic length $\ell_B$) for adding (removing) a magnetic flux at $\nu=1/2$.
  Although the finite size effect are large, it is clearly seen that the energy needed to add (remove) a magnetic flux is negative (positive).
  See comparison with $\nu=1/3$ in inset.
  The dashed lines show the extrapolation of the gap to the thermodynamic limit.
  \\
  {\bf Inset:}
  The charge gap at constant torus area $A$ for adding (removing) a magnetic flux at $\nu=1/3$, as comparison.
  Here there is a (positive) energy gap both for adding or removing a magnetic flux.
  }
  \label{fig:Charge_gap_A}
\end{figure}

In Fig.~\ref{fig:Charge_gap_A} we compute the charge gap for the the $\nu_\pm$ states. 
Although there are clearly visible finite size effects, the data appear to support that the charge gap for removing a flux in the thermodynamics limit is positive $\Delta_{-}\approx 0.05$ $\frac{e^2}{\epsilon_0\ell_B}$, and for adding one flux is negative $\Delta_{+}\approx -0.02$ $\frac{e^2}{\epsilon_0\ell_B}$.
For comparison we may consider the charge gap at $\nu=1/3$ (see inset in Fig.~\ref{fig:Charge_gap_A}), 
where there is an energy cost both for removing and adding magnetic flux.
Note that the jagged structure in the charge gap is due to the irregular energy of the global ground state at half filling, rather than effect coming from the $\nu_\pm$ ground states.
Also note that the second order difference measure for the stability $E_++E_--2E_0=\Delta_++\Delta_-$ gives a result (in the thermodynamic limit) that is independent of whether the magnetic length or area is held constant.

\section{Linear vs. Quadratic Dispersion}
\label{sec:linear_vs_quadratic}
Here we comment on the discussion on the true dispersion relation of the composite fermions.
If the CFs are Dirac Fermions they should in principle have linear dispersion at least at small $k$,
although this is not necessary at larger $k$.
To directly probe the dispersion of the composite fermions we may attempt to fit the CF-energy defined in \eqref{eq:cf_energy} to the actual energy of the lowest energy state in each momentum sector.
We attempt to fit the ground state energy in all sectors using a simple model of the dispersion \eg in the form 
\begin{equation}
E_{k}=E_{0}+c_{\alpha}\left(\ell_B\left|k\right|\sin\alpha+\ell_B^2\left|k\right|^{2}\cos\alpha\right)\label{eq:CF_Energy_Model}
\end{equation}
 where $\alpha$ tunes between a quadratic and a linear dispersion.
We note that stable Fermi disc would exist in the range $-\frac{\pi}{2}<\alpha\leq\frac{\pi}{2}$,
although to avoid Mexican hat potentials we should restrict $0\leq\alpha\leq\frac{\pi}{2}$.

To determine the preferred value of $E_0$, $c_\alpha$ and $\alpha$ we scan over $\alpha$ and minimize using least squares fit for the parameters $E_0$ and $c_\alpha$.
Unfortunately we find that the CF energy landscape in many cases is quite insensitive to the exact value of $\alpha$ and quite different $\alpha$ can still generate qualitatively similar energy landscapes.
This should probably not be too surprising as for large $N_e$ one may linearize around the Fermi momentum $k_F$ and obtain the energy $E_k\approx \tilde E_0+\tilde c_\alpha \delta_k$
where $\tilde E_0=E_0+c_{\alpha}k_{F}\left(\sin\alpha+\cos\alpha\right)$,
$\tilde c_\alpha=c_\alpha\left(\sin\alpha+2k_F\ell\cos\alpha\right)$ and
$\delta_k=|k|-k_F$.
The Fermi momentum is $k_F=\frac{2\pi n_F}{L}=\frac{2\pi \sqrt{\frac{N_e}\pi}}{\sqrt{2\pi N_s}\ell_B}=\sqrt{\frac{2N_e}{N_s}}\frac1{\ell_B}=\frac1{\ell_B}$.

\begin{figure}
\begin{centering}
\includegraphics[width=1\linewidth]{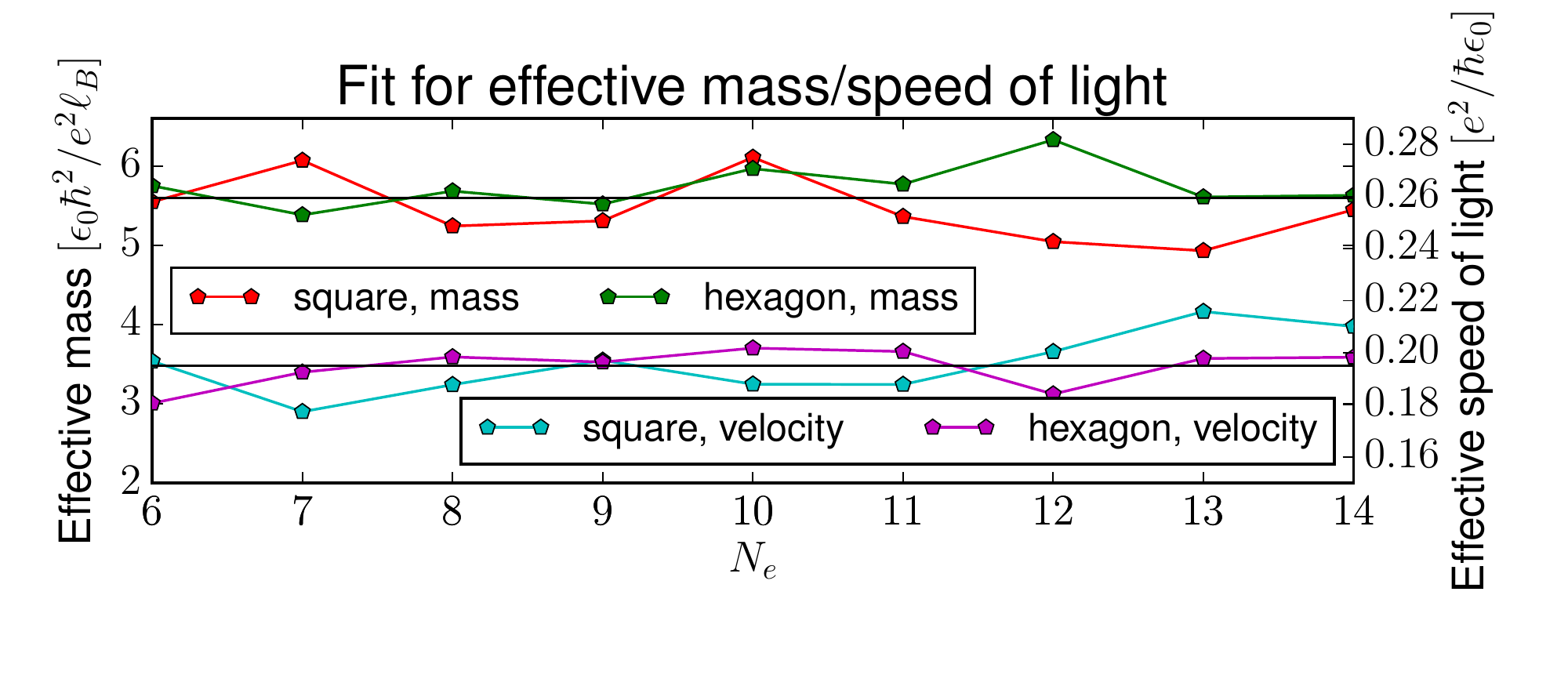}
\par\end{centering}
\caption{The estimate of the effective mass for non-relativistic CFs or the speed off light  for Dirac CFs for a square and hexagonal torus. We find that the masses and velocities are stable in a wide range of $N_e$ and for both square and hexagonal tori. \label{fig:CF_effective_mass}}
\end{figure}

Although we cannot determine which (if any) of $\alpha=0$  (quadratic) or $\alpha=\frac{\pi}2$  (linear) is preferred we can still determine the effective mass $m^\star$ or effective speed of light $c^\star$ that the corresponding composite fermions would have if their dispersion actually were one of the above.
By equating \eqref{eq:CF_Energy_Model} with either $c^\star\hbar|k|$ or $\frac{\hbar^2|k|^2}{2m^\star}$ we find that $c^\star = c_{\frac{\pi}2}\ell_B/\hbar$ and $m^\star = \hbar^2/(2\ell_B^2 c_{0})$.
We find that $c_0=0.090\pm0.006$ $\frac{e^2}{\epsilon_0}$ and $c_{\frac{\pi}2}=0.195\pm0.009$  $\frac{e^2}{\epsilon_0}$ quite independently of system size.
The mass and velocity come out roughly the same for square and hexagonal tori, and are consistent with the observation that $c^\star m^\star\approx\hbar/\ell_B$ when linearizing around the Fermi momentum.
This translates into $m^\star=5.6\pm0.2$ $\epsilon_0\hbar^2/(e^2\ell_B)$ and $c^\star=0.195\pm0.010$ $e^2/(\epsilon_0\hbar)$, see Figure \ref{fig:CF_effective_mass}.

Our result is close to other values for the effective mass $m^\star=6\epsilon_0\hbar^2/(e^2\ell_B)$  and $m^\star=5\epsilon_0\hbar^2/(e^2\ell_B)$ reported in the literature in Ref.~\onlinecite{Yu98} and \onlinecite{Morf95} respectively.

\section{Berry Phase}\label{sec:Berry_phase}
\newcommand{\Tikzscale}{0.7}
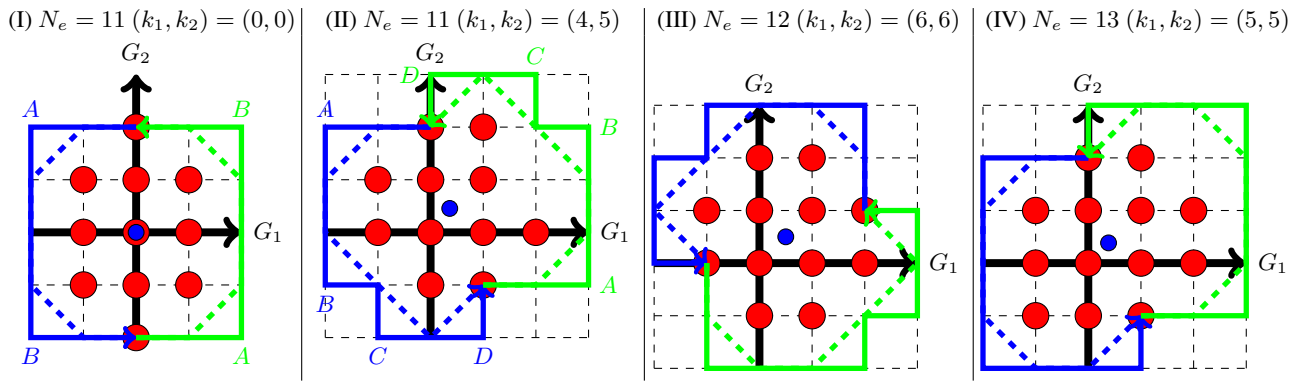
\begin{figure*}
  \begin{center}
  \begin{tabular}{c|c|c|c}
    (I) $N_e=11$ $(k_1,k_2)=(0,0)$ &
    (II) $N_e=11$ $(k_1,k_2)=(4,5)$ &
    (III) $N_e=12$ $(k_1,k_2)=(6,6)$ &
    (IV) $N_e=13$ $(k_1,k_2)=(5,5)$ \\
  \begin{tikzpicture}[scale=\Tikzscale]
    \coordinate (CM) at (0,0);
    \coordinate (E1) at (1,-1);
    \coordinate (E2) at (1,0);
    \coordinate (E3) at (1,1);
    \coordinate (E4) at (0,-1);
    \coordinate (E5) at (0,0);
    \coordinate (E6) at (0,1);
    \coordinate (E7) at (-1,-1);
    \coordinate (E8) at (-1,0);
    \coordinate (E9) at (-1,1);
    \coordinate (E10) at (0,2);
    \coordinate (E11) at (0,-2);
    \foreach \x in {-2,...,2}
             {\draw[dashed] (\x,-2) --  (\x,2);
               \draw[dashed] (-2,\x) --  (2,\x);}
             \draw[line width=3pt,->] (0,-2) --  (0,3) node[above]{$G_2$};
             \draw[line width=3pt,->] (-2,0) --  (2,0) node[right]{$G_1$};
    \draw[fill=red] (E1) circle (.25) ;
    \draw[fill=red] (E2) circle (.25) ;
    \draw[fill=red] (E3) circle (.25) ;
    \draw[fill=red] (E4) circle (.25) ;
    \draw[fill=red] (E5) circle (.25) ;
    \draw[fill=red] (E6) circle (.25) ;
    \draw[fill=red] (E7) circle (.25) ;
    \draw[fill=red] (E8) circle (.25) ;
    \draw[fill=red] (E9) circle (.25) ;
    \draw[fill=red] (E10) circle (.25) ;
    \draw[fill=red] (E11) circle (.25) ;
    \draw[fill=blue] (CM) circle (.15) ;

    \draw[->,blue,line width=2pt] (0,2) -- (-2,2) node[above]{$A$} -- (-2,-2) node[below]{$B$}  -- (0,-2);
    \draw[dashed,->,blue,line width=2pt] (0,2) -- (-1,2) --(-2,1) -- (-2,-1) -- (-1,-2) -- (0,-2);
    \draw[->,green,line width=2pt] (0,-2) -- (2,-2) node[below]{$A$} -- (2,2) node[above]{$B$} -- (-0,2);
    \draw[dashed, ->,green,line width=2pt] (0,-2) -- (1,-2) -- (2,-1) -- (2,1) --(1,2) -- (-0,2);
  \end{tikzpicture}
  &
  \begin{tikzpicture}[scale=\Tikzscale]
    \coordinate (CM) at (4/11,5/11);
    \coordinate (E1) at (0,0);
    \coordinate (E2) at (1,0);
    \coordinate (E3) at (0,1);
    \coordinate (E4) at (0,-1);
    \coordinate (E5) at (-1,0);
    \coordinate (E6) at (1,1);
    \coordinate (E7) at (1,-1);
    \coordinate (E8) at (-1,1);
    \coordinate (E9) at (0,2);
    \coordinate (E10) at (1,2);
    \coordinate (E11) at (2,0);
    \foreach \x in {-2,...,3}
             {\draw[dashed] (\x,-2) --  (\x,3);
               \draw[dashed] (-2,\x) --  (3,\x);}
             \draw[line width=3pt,->] (0,-2) --  (0,3) node[above]{$G_2$};
             \draw[line width=3pt,->] (-2,0) --  (3,0) node[right]{$G_1$};
    \draw[fill=red] (E1) circle (.25) ;
    \draw[fill=red] (E2) circle (.25) ;
    \draw[fill=red] (E3) circle (.25) ;
    \draw[fill=red] (E4) circle (.25) ;
    \draw[fill=red] (E5) circle (.25) ;
    \draw[fill=red] (E6) circle (.25) ;
    \draw[fill=red] (E7) circle (.25) ;
    \draw[fill=red] (E8) circle (.25) ;
    \draw[fill=red] (E9) circle (.25) ;
    \draw[fill=red] (E10) circle (.25) ;
    \draw[fill=red] (E11) circle (.25) ;
    \draw[fill=blue] (CM) circle (.15) ;
    \draw[->,blue,line width=2pt] (0,2) -- (-2,2)  node[above]{$A$} -- (-2,-1)  node[below]{$B$} -- (-1,-1) -- (-1,-2) node[below]{$C$} -- (1,-2)  node[below]{$D$} -- (1,-1) ;
    \draw[dashed,->,blue,line width=2pt] (0,2) -- (-1,2) --(-2,1) -- (-2,0) -- (0,-2) -- (1,-1) ;
    \draw[->,green,line width=2pt] (1,-1) -- (3,-1) node[right]{$A$} -- (3,2)  node[right]{$B$} -- (2,2) -- (2,3) node[above]{$C$} -- (0,3) node[left]{$D$} -- (0,2);
    \draw[dashed,->,green,line width=2pt] (1,-1) -- (2,-1) -- (3,0) -- (3,1) -- (1,3) -- (0,2);
  \end{tikzpicture}
  &
  \begin{tikzpicture}[scale=\Tikzscale]
    \coordinate (CM) at (.5,.5);
    \coordinate (E1) at (0,0);
    \coordinate (E2) at (1,0);
    \coordinate (E3) at (0,1);
    \coordinate (E4) at (0,-1);
    \coordinate (E5) at (-1,0);
    \coordinate (E6) at (1,1);
    \coordinate (E7) at (1,-1);
    \coordinate (E8) at (-1,1);
    \coordinate (E9) at (0,2);
    \coordinate (E10) at (1,2);
    \coordinate (E11) at (2,0);
    \coordinate (E12) at (2,1);
    \foreach \x in {-2,...,3}
             {\draw[dashed] (\x,-2) --  (\x,3);
               \draw[dashed] (-2,\x) --  (3,\x);}
             \draw[line width=3pt,->] (0,-2) --  (0,3) node[above]{$G_2$};
             \draw[line width=3pt,->] (-2,0) --  (3,0) node[right]{$G_1$};
    \draw[fill=red] (E1) circle (.25) ;
    \draw[fill=red] (E2) circle (.25) ;
    \draw[fill=red] (E3) circle (.25) ;
    \draw[fill=red] (E4) circle (.25) ;
    \draw[fill=red] (E5) circle (.25) ;
    \draw[fill=red] (E6) circle (.25) ;
    \draw[fill=red] (E7) circle (.25) ;
    \draw[fill=red] (E8) circle (.25) ;
    \draw[fill=red] (E9) circle (.25) ;
    \draw[fill=red] (E10) circle (.25) ;
    \draw[fill=red] (E11) circle (.25) ;
    \draw[fill=red] (E12) circle (.25) ;
    \draw[fill=blue] (CM) circle (.15) ;

    \draw[->,blue,line width=2pt] (2,1) -- (2,2) -- (2,3) -- (1,3) -- (0,3) -- (-1,3) -- (-1,2) -- (-2,2) -- (-2,1) -- (-2,0) -- (-1,0);
    \draw[dashed,->,blue,line width=2pt] (2,1) -- (2,2) -- (1,3) -- (0,3) -- (-2,1) --  (-1,0);
    \draw[->,green,line width=2pt] (-1,0) -- (-1,-2) -- (2,-2) -- (2,-1) -- (3,-1) -- (3,1) -- (2,1);
    \draw[dashed,->,green,line width=2pt] (-1,0) -- (-1,-1) -- (0,-2) -- (1,-2) -- (3,0) -- (2,1);
  \end{tikzpicture}
  &
  \begin{tikzpicture}[scale=\Tikzscale]
    \coordinate (CM) at (5/13,5/13);
    \coordinate (E1) at (0,0);
    \coordinate (E2) at (1,0);
    \coordinate (E3) at (0,1);
    \coordinate (E4) at (0,-1);
    \coordinate (E5) at (-1,0);
    \coordinate (E6) at (1,1);
    \coordinate (E7) at (1,-1);
    \coordinate (E8) at (-1,1);
    \coordinate (E9) at (0,2);
    \coordinate (E10) at (1,2);
    \coordinate (E11) at (2,0);
    \coordinate (E12) at (2,1);
    \coordinate (E13) at (-1,-1);
    \foreach \x in {-2,...,3}
             {\draw[dashed] (\x,-2) --  (\x,3);
               \draw[dashed] (-2,\x) --  (3,\x);}
             \draw[line width=3pt,->] (0,-2) --  (0,3) node[above]{$G_2$};
             \draw[line width=3pt,->] (-2,0) --  (3,0) node[right]{$G_1$};
    \draw[fill=red] (E1) circle (.25) ;
    \draw[fill=red] (E2) circle (.25) ;
    \draw[fill=red] (E3) circle (.25) ;
    \draw[fill=red] (E4) circle (.25) ;
    \draw[fill=red] (E5) circle (.25) ;
    \draw[fill=red] (E6) circle (.25) ;
    \draw[fill=red] (E7) circle (.25) ;
    \draw[fill=red] (E8) circle (.25) ;
    \draw[fill=red] (E9) circle (.25) ;
    \draw[fill=red] (E10) circle (.25) ;
    \draw[fill=red] (E11) circle (.25) ;
    \draw[fill=red] (E12) circle (.25) ;
    \draw[fill=red] (E13) circle (.25) ;
    \draw[fill=blue] (CM) circle (.15) ;

    \draw[->,blue,line width=2pt] (0,2) -- (-2,2) -- (-2,-2) -- (1,-2) -- (1,-1);
    \draw[dashed,->,blue,line width=2pt] (0,2) -- (-1,2) -- (-2,1) -- (-2,-1)--(-1,-2)--(0,-2) -- (1,-1);
    \draw[->,green,line width=2pt] (1,-1) -- (3,-1) -- (3,3) -- (0,3) -- (0,2);
    \draw[dashed,->,green,line width=2pt] (1,-1) -- (2,-1) -- (3,0) -- (3,2) -- (2,3)--(1,3)--(0,2);
  \end{tikzpicture}
  \end{tabular}
  \end{center}
  \caption{Path of dragging two CFs (red discs) aground the Fermi disc such that momentum is conserved. The two colors (blue,green) show the paths for the two CFs on either side of the Fermi disc.
    After the move, the Fermi disc is returned to its original configuration.
    There is a phase of $\pi$ from the exchange of the electrons and in addition a Berry phase. \label{fig:CF_Berry_Pahse}}
  \end{figure*}

It has recently been proposed in the literature that if the CFs are Dirac fermions then they should yield a Berry phase equal to $\pi$ if a single fermion is dragged around the Fermi disc.   This phase occurs at tree level in the Dirac picture\cite{Son2015}, but not at mean field level in the Halperin-Lee-Read picture\cite{Halperin93} (although the phase should reappear once corrections are included\cite{Wang2017}).   

On a plane, for non-interacting fermions, dragging a fermion around the Fermi surface is the same as continuously changing its momentum $\mathbf{k}$ in a loop.
For CFs, the whole picture is complicated by the fact that we do not have single particle CF wave functions in the conventional sense.
Rather the momentum $\mathbf k$ is dressed with a Jastrow factor (see \eqref{eq:CF_wf}) when the CFs bind two flux quanta.
Thus dragging $\mathbf{k}$ in a loop means to change the $\mathbf{k}$ in the exponential, an operation that affects the entire many-body wave function.

On the torus the matter is further complicated by the available momenta being discrete and wave functions with different total momenta $\mathbf{K}$ being orthogonal to each other.
Thus, changing $\mathbf{k}^j$ in discrete steps around the Fermi disc, is singular in the sense that the obtained Berry phase if this procedure would be $e^{\i\theta}=0$ identically.
We may however move two CFs at opposite sides of the Fermi disc at the same time while preserving the total momentum.
When the two CFs have performed half a revolution of $N$ steps, they have collectively performed a full revolution around the Fermi disc of $2N$ steps.

\begin{table}
\begin{tabular}{|c|c|c|c|c|}
\hline
 Case & I & II & III & IV\tabularnewline
\hline
\hline
$N=$ Steps in full path & 8 & 10 & 10 & 10\tabularnewline
\hline
No corners  Cut& $6.94\pi$ & $8.71\pi$ & $8.84\pi$ & $8.89\pi$\tabularnewline
\hline
Cut corner $A$  ( -1 step) & $6.02\pi$ & $7.83\pi$ & $7.91\pi$ & $7.87\pi$\tabularnewline
\hline
Cut corner $B$ ( -1 step) & $6.00\pi$ & $7.79\pi$ & $7.99\pi$ & $7.95\pi$\tabularnewline
\hline
Cur corner $A$ \& $B$  ( -2 steps)& $5.08\pi$ & $6.91\pi$ & $7.07\pi$ & $6.93\pi$\tabularnewline
\hline

\end{tabular}
\caption{The Berry phase accumulated when stepping through $N$ FS states to let two composite fermions trade places.
  The four considered paths are show in Fig.~\ref{fig:CF_Berry_Pahse}.
  We consider moving two particles along the solids paths in the figure, and compute the Berry phase that is accumulated.
  We also consider cutting the corners at $A$ and/or $B$ (dashed lines in Fig.~\ref{fig:CF_Berry_Pahse}) and repeat the procedure.
  We find that when a corner is cut the Berry phase is reduced by $\sim\pi$ and when two cornets are cut, then Berry phase is reduced by $\sim2\pi$ compared with the path with no corner cut.
  The phase that is accumulated is approximately $e^{\i\theta}=(-1)^{N-1}$, but there are finite size effects, and the phase we measure differs from the expected phase by typically $0.1\pi$, although there are cases in the table with errors as much as $0.3\pi$.}
\label{tab:Berry_Pahse}
\end{table}

As the different FS states have low overlap with each other along the path, the Berry phase is at best ill conditioned.
We remedy this by constructing a set of intermediate states that continuously connects the different FS-states. The intermediate states are constructed as the lowest eigenvalue of the projector
\[ P_{a,l}=1-(1-a)\ketbra{k_l}{k_l}-a\ketbra{k_{l+1}}{k_{l+1}}.\]
We can then use the formula 
\begin{equation}
  e^{\i\theta}\approx\prod_{l=0}^{N-1}\braket{k_l}{k_{l+1}},\label{eq:Berry_pahse}
  \end{equation}
to evaluate the Berry phase.

We perform the Berry phase calculation for a selection of Fermi discs at
(I) $N_e=11$ with $\mathbf{K}=(0,0)$ as well well as the
the ground states at (II) $N_e=11$ with $\mathbf K=(4,5)$,
the ground states at (III) $N_e=12$ with $\mathbf K=(6,6)$ and $N_e=13$ with (IV) $\mathbf K=(5,5)$.
The paths of these exchanges can be seen in Fig.~\ref{fig:CF_Berry_Pahse}.

As two CFs have traded place while circumnavigating the Fermi Disc the last and first state differ my a minus sign $\ket{k_0}=-\ket{k_N}$. The Berry phases that we obtain are reported in Tab. \ref{tab:Berry_Pahse} and do no contain this extra minus sign due to the exchange.

For path (I) we note that the states at position $A$ and $B$ are high energy states,
and we may attempt to ignore them by cutting the corners at $A$ and/or $B$.
In doing so we find that cutting a corner (in all four paths considered) reduces the overall phase by $\pi$.
Thus by cutting both the corners at $A$ and $B$ we reduce the Berry phase of $2\pi$.
The full Berry phase that we acquire is thus $e^{\i \theta}\sim e^{\i \pi (N-1)}$, where $N$ counts the number of FS states in the path.
There are finite size effects, and the phase we measures differs from the expected phase by typically $0.1\pi$, although there are cases in the table with errors as much as $0.3\pi$.

We have recently become aware that Wang et.~al.\cite{Wang2017}~and Geraedts et.~al\cite{Geraedts_2017_Berry}~also have performed a calculation of this type. They use a modified wave function that allows them to circumvent the projection problem  and target much larger system sizes than is amenable to ED.
In their work, they find that the total Berry phase is $e^{\i\theta}=\i^{N_+}(-\i)^{N_-}(-1)^W$
where $W$ is the winding number, while $N_\pm$ is the number of (anti-)clockwise steps through FS-states that have been taken in the path.
Their results are consistent with what we see in our numerics, since as we are moving two particles (half a revolution) we should see $e^{\i\theta}=(\i^{N}(-\i)^0(-1)^{\frac12})^2=(-1)^{N-1}$ according to their formula.
We clearly see in our numerics the $N$ dependence in $(-1)^{N-1}$, however our system sizes seem to small to be able to fully resolve the phase $(-1)^W$ that Wang et.~al. reports.

Lastly we remark that as the wave functions used in Ref.~\onlinecite{Wang2017} are not the result of a controlled projection it would be interesting to study how those wave functions differ from the exactly projected wave functions used in this paper.

\section{Discussion and Outlook}\label{sec:Discussion}
In this paper we have systematically tested the composite fermion wave functions at filling fraction $\nu=1/2$.
We find that the trial wave functions give an excellent description of the global Coulomb ground state, but also do very well in modeling most of lowest energy states in all momentum sectors.
The CF wave functions also model the charged excitations at $\nu_\pm=\frac{N_{e}}{2N_{e}\pm1}$ to high accuracy.
An interesting question is how the CF state at $\nu_\pm$ is connected to the state at half filling.
It would be nice to better understand if/how the picture of filling an increasing number of -- more closely packed --  $\Lambda$-levels connects to the filled CF Fermi disc.

We also find that the low energy CF states are almost fully particle hole symmetric, but that particle hole symmetry is broken at higher energies, especially on the hexagonal torus.

The methods used in this paper can be used to project any wave function written in real space to the LLL, and it would be interesting to compare the exactly projected wave functions with the wave functions developed in Ref.~\onlinecite{Wang2017} and Ref.~\onlinecite{Pu2017}.

The work presented here is in agreement with the work in Refs.~\onlinecite{Wang2017} and \onlinecite{Geraedts_2017_Berry} showing that the Berry phase picked up when a composite fermion is moved around the Fermi surface is $\pi$.
There seems however to be a contribution to the Berry phase that depends on the number of FS-configurations that that $\mathbf{k}$-space path traverses.
This extra phase is not accounted for in the simplest of the Dirac composite fermions descriptions and does not seem to be possible to explain away as an Aharonov Bohm-phase.
All in all, this warrants more investigations into the nature of the composite fermions at half filling.
The Berry phase calculation would be interesting to carry out in the plane as it would allow for determining the local Berry curvature, and its properties, by taking advantage of the continuous set of momenta.

\section{Acknowledgements}
We thank Nicolas Regnault, Cecile Repellin, Ajit C Balram, Jainendra Jain and Lars Fritz for discussions. This work was supported through SFI Principal Investigator Award 12/IA/1697. We also wish to acknowledge the SFI/HEA Irish Centre for High-End Computing (ICHEC) for the provision of computational facilities and support.  SHS is supported by EPSRC grants EP/I031014/1 and EP/N01930X/1. Statement of
compliance with EPSRC policy framework on research
data: This publication is theoretical work that does not
require supporting research data.

\bibliographystyle{h-physrev3} 
\bibliography{CF_Sea}

\appendix

\section{Exact LLL projection of the CF wave functions}
\label{sec:LLL_proj}
The exact projection of formula \eqref{eq:CF_projected} of \eqref{eq:CF_wf} has been known for many years in the literature\cite{Read98} and can be argued for starting from Ref. \onlinecite{Girvin84}.
The formula does however not seem to have a published derivation anywhere, so a service to the reader, we provide that.
We begin by noting that we can write the composite Fermi liquid at
$\nu=\frac{1}{2}$ as
\[
\PsiCFL=P_{\mathrm{LLL}} \mathcal{A}\left[e^{\i2\pi\sum_{i=1}^{N_e}\left(k^i_xx_i+k^i_yy_i\right)}\psi_{\frac{1}{2}}\right],
\]
 where $\psi_{\frac{1}{2}}$ is the bosonic Laughlin state $\nu=\frac{1}{2}$
and  $k_x^i$, $k_y^i$ the labels the collection of occupied CF-momenta $\mathbf{k}=(k_x,k_y)$, and $z_j=L(x_j+\tau y_j)$.
For notational simplicity we will drop the index $i$ and the sum $\sum_{i=1}^{N_e}$, such that
\eg $k_xx\equiv \sum_{i=1}^{N_e}=k^i_xx_i$.
We can do this as the projections for the different electrons commute.

We first note that since $k_x,k_y\in\mathbb{Z}$, the exponential factor
$e^{\i2\pi\left(k_xx+k_yy\right)}$ is periodic in $x$ and $y$ (with period 1).
As a consequence, any gauge dependence only affects $\psi_{\frac12}$.
We can therefore, without loss of generality, choose to work in symmetric gauge.

The ladder operators in symmetric gauge are
\[
  a^{\dagger} = \sqrt{2}\left(\frac{1}{4}\bar{z}-\partial_{z}\right)
  ,\quad\quad
  b^{\dagger} = \sqrt{2}\left(\frac{1}{4}z-\partial_{\bar{z}}\right)
 \]
where $a^\dagger$ increases the Landau level index and $b^\dagger$ increases the angular momenta within a LL.
We can express $z,\bar{z},\partial_{z},\partial_{\bar{z}}$ in terms of $a,a^{\dagger},b,b^{\dagger}$ and also by extension $x$ and $y$. This gives 
\begin{eqnarray}
y 
 & = & \frac{a+b^{\dagger}-a^{\dagger}-b}{\i\sqrt{2}L\tau_{2}}\nonumber\\
x 
 & = & \frac{\left(a^{\dagger}+b\right)\tau-\left(a+b^{\dagger}\right)\bar{\tau}}{\i\sqrt{2}L\tau_{2}}.\label{eq:ap_x_y}
\end{eqnarray}
 Inserting \eqref{eq:ap_x_y} into the exponential gives
\[ k_xx+k_yy =
 \frac{1}{\i\sqrt{2}}\left[\bar{k}a-ka^{\dagger}\right]+\frac{1}{\i\sqrt{2}}\left[\bar{k}b^{\dagger}-kb\right]\]
 where
\begin{eqnarray}
  \bar{k} = \frac{1}{L\tau_{2}}\left(k_{y}-k_{x}\bar{\tau}\right)
  &,\quad\quad&
 k = \frac{1}{L\tau_{2}}\left(k_{y}-k_{x}\tau\right).\label{eq:app_k_k_bar}
\end{eqnarray}
 We can now write $\PsiCFL$ as
\[
\PsiCFL=P_{\mathrm{LLL}} \mathcal{A}\left[e^{\sqrt{2}\pi\left(\bar{k}a-ka^{\dagger}\right)}e^{\sqrt{2}\pi\left(\bar{k}b^{\dagger}-kb\right)}\psi_{\frac{1}{2}}\right],
\]
 and since $\left[P_{\mathrm{LLL}} ,b\right]=\left[P_{\mathrm{LLL}} ,b^{\dagger}\right]=0$ we can pull the $b$ operators to the left and we have
\[
\PsiCFL=\mathcal{A}\left[e^{\sqrt{2}\pi\left(\bar{k}b^{\dagger}-kb\right)} P_{\mathrm{LLL}}
e^{\sqrt{2}\pi\left(\bar{k}a-ka^{\dagger}\right)}\psi_{\frac{1}{2}}\right].
\]
Next, using the CBH-formula $e^{va^{\dagger}-\bar{v}a}=e^{-\frac{1}{2}\left|v\right|^{2}}e^{va^{\dagger}}e^{-\bar{v}a}$,
we have
\begin{eqnarray*}
  \PsiCFL & = & \mathcal{A} \left[e^{\sqrt{2}\pi\left(\bar{k}b^{\dagger}-kb\right)} P_{\mathrm{LLL}}
    e^{-2\pi^{2}\left|k\right|^{2}}e^{-\sqrt{2}\pi ka^{\dagger}}e^{\sqrt{2}\pi\bar{k}a}\psi_{\frac{1}{2}} \right]\\
  & = & \mathcal{A} \left[e^{\sqrt{2}\pi\left(\bar{k}b^{\dagger}-kb\right)} P_{\mathrm{LLL}}
    e^{-2\pi^{2}\left|k\right|^{2}}e^{-\sqrt{2}\pi ka^{\dagger}}\psi_{\frac{1}{2}} \right]\\
  & = & \mathcal{A} \left[e^{\sqrt{2}\pi\left(\bar{k}b^{\dagger}-kb\right)}
    e^{-2\pi^{2}\left|k\right|^{2}}\psi_{\frac{1}{2}}\right]\\
    & = & e^{-2\pi^2\sum_j\left|k_j\right|^{2}}\mathcal{A} \left[e^{\sqrt{2}\pi\left(\bar{k}b^{\dagger}-kb\right)}
    \psi_{\frac{1}{2}}\right].
\end{eqnarray*}
 The second line is due to $a_j \psi_{\frac{1}{2}}=0$ for all $j$,
the third is since $P_{\mathrm{LLL}} \left(a^{\dagger}\right)^{n}\psi_{\frac{1}{2}}=0$
if $n>0$ for all $j$.
In the last line we pull $\exp\left(-2\pi^2\left|k\right|^{2}\right)$ out of the antisymmetrization.

Inspecting $e^{\sqrt{2}\pi\left[\bar{k}b^{\dagger}-kb\right]}$,
we see that this can be interpret as magnetic translation of $t_{-k_y,k_x}$ acting on $z$.
To see this clearly we write
\begin{eqnarray*}
  b & = &
  =\frac{\sqrt{2}}{4}L\left(x+\bar{\tau}y\right)
  +\frac{\sqrt{2}\left(-\bar{\tau}\partial_{x}+\partial_{y}\right)}{L\left(\tau-\bar{\tau}\right)}\\
  b^{\dagger} & = &
  =\frac{\sqrt{2}}{4}L\left(x+\tau y\right)
  +\frac{\sqrt{2}\left(-\tau\partial_{x}+\partial_{y}\right)}{L\left(\tau-\bar{\tau}\right)}.
\end{eqnarray*}
This gives (up to terms linear in $x$ and $y$ - which are gauge dependent)
\[
\sqrt{2}\pi\left[\bar{k}b^{\dagger}-kb\right] =-\frac{k_{y}}{N_{\phi}}\partial_{x}+\frac{k_{x}}{N_{\phi}}\partial_{y}+\mathcal{O}(x)+\mathcal{O}(y).
\]
We can thus write the state $\PsiCFL$ as
\begin{eqnarray}
\PsiCFL & = & e^{-2\pi^{2}\sum_{j}\left|k_{j}\right|^{2}}\mathcal{A}\left[\prod_j t_{-k^j_y,k^j_x}\psi_{\frac{1}{2}}\right].\label{eq:Final_eqn}
\end{eqnarray}
Using \eqref{eq:app_k_k_bar} we may also express $\left|k\right|^{2}$ as
\begin{eqnarray*}
\left|k\right|^{2} & = & \frac{1}{L^2\tau_{2}^2}\left|k_{y}-k_{x}\tau\right|^2\\
 & = & \frac{k_{y}^{2}-2k_{y}k_{x}\tau_{1}+k_{x}^{2}\left|\tau\right|^{2}}{2\pi N_{s}\ell_B^2\tau_{2}}.
\end{eqnarray*}

\section{Algorithm to find the lowest pseudo-energy FS-states}
\label{sec:Algorithm}
A state with $N$ CFs in distinct CF orbitals with momenta $\mathbf{k}_{i}$ has pseudo-energy $E=\sum_{i}\epsilon_{i}$. Here, the single particle pseudo-energies $\epsilon_{i}$ depend on both $\mathbf{k}_{i}$ and the total momentum $\mathbf{K}$, we have $\epsilon_{i}=\epsilon(\mathbf{k}_{i}-\mathbf{K}/N)$.
We now present an algorithm which serves to determine the $M$ sets $\{k_{i}\}$ which have the lowest pseudo-energy.  

First of all we choose $\mathbf{K}$ and order the single particle momenta $\mathbf{k}_{i}$ by their pseudo-energies $\epsilon_{i}$, so $\epsilon_1\le\epsilon_2\le \epsilon_3\le...$. (If some of the $\epsilon_{i}$ are equal we order them however we like.)

We define $I^{m}=\{i^{m}_{1},i^{m}_{2},...,i^{m}_{N}\}$ to be the ordered list containing the indices of the set of $N$ single particle momenta so that the total pseudo-energy $E(\mathbf{K})$ takes its $m^{\mathrm{th}}$ lowest value. Here we consider all possible sets of momenta, so for example $I^{0}=\{1,2,...,N\}$, regardless whether $\sum_{i}\mathbf{k}_{i}=\mathbf{K}$.  There may be some ambiguity in the definition of $I^{m}$ if there are sets of momenta with exactly the same total pseudo-energy. This has no impact on the algorithm. We simply require that $E(I^1)\le E(I^2)\le E(I^3)\le...$. 
We will collect the index sets for the $M$ lowest configurations satisfying $\sum_{i}\mathbf{k}_{i}=\mathbf{K}$ in an ordered list $T$ (again there can be some ambiguity in the order if there are sets that yield the same energy). In the algorithm we also make use of a further list $L$ of index sets ordered by total pseudo-energy. We now proceed as follows

\begin{description}
\item{Initialization:}\\
Set $L=\{I^{0}\}$. \\
If $\sum_{i=1}^{N}\mathbf{k}_{i}=\mathbf{K}$, set $T=L$.  Otherwise set $T=\emptyset$. 
\item{Iteration:}\\
Given $I^{m}$, we formally define  $i^{m}_{N+1}=N^2+1$ and consider all $p$ such that $i_{p+1}>i_{p}+1$.  These are the indices for which the orbital after the $p^{\mathrm{th}}$ orbital is empty in the state described by $I^{m}$. Construct the index sets $I^{m,p}=\{i^{m,p}_{1},i^{m,p}_{2},...,i^{m,p}_{N}\}$ given by $i^{m,p}_{l}=\delta_{p,l}( i^{m}_{p}+1)+(1-\delta_{p,l}) i^{m}_{l}$. These sets describe states in which the particle in the $p^{\mathrm{th}}$ orbital has been moved to the next orbital in the pseudo-energy ordering. The reason for the definition of  $i^{m}_{N+1}$ is to make sure that the particle in the $N^{\mathrm{th}}$ orbital can always be moved, unless this orbital is the highest energy orbital (in which case $i^{m}_{N}=N^2$). \\ Add those  $I^{m,p}$ which are not already contained in the ordered list $L$ to $L$ (making sure $L$ remains ordered by pseudo-energy). \\ Set $I^{m+1}$ to be the first (lowest pseudo-energy) element of the list $L$. (There may be multiple elements of $L$ with this energy but this does not matter to us here).\\ Remove $I^{m+1}$ from $L$.  \\If  $\mathbf{K}(I^{m+1})=\mathbf{K}$ then add $I^{m+1}$ tot the list $T$. 
\end{description}
We repeat the iteration step until the list $T$ has the desired length $M$. At this point there may still be further index sets which lead to the same pseudo-energy as the last set in $T$, but no index sets with lower pseudo-energy exist. Note that the list $L$ in this algorithm grows at most linearly with the number of iterations, but in reality it will almost always grow more slowly at low energy, as it typically only contains states near the current Fermi ``circle". In other words, as long as $I^{m}$ represents a fairly compact Fermi ``disc" then we expect the length of $L$ to be of the order of the circumference of this ``disc".  Also with reasonably isotropic pseudo-dispersion, we expect that we should find a state with the desired momentum roughly every $N^2$ iterations, since there are $N^2$ momentum sectors, so we can naively estimate that the algorithm will take about $MN^2$ iterations to find the desired states.

\section{One or two composite fermion Fermi liquids?}
\label{sec:Laughlintorus_app}

The wave function for the composite Fermi liquid is not completely specified by Eq.~\eqref{eq:CF_projected} as the bosonic Laughlin ground state wave function which appears, as $\psi_{\frac{1}{2}}$ is not unique. On the torus, there is a degenerate doublet of ground states, with a basis for the two dimensional  space given by momentum eigenstates at $\mathbf{K}=(0,0)$ and  $\mathbf{K}=(N_e,0)$. We can write these wave functions explicitly as 
\begin{eqnarray}
  \psi_{\frac{1}{2}}^{(s)}&=&e^{-\frac{1}{2}\sum_{i}y_{i}^{2}} \prod_{i<j}\elliptic 1{z_{ij}}{\tau}^{2}
  \\
  &\times&\ellipticgeneralized{\frac sq+\frac{1}{2}(N_e-1)}{\frac{1}{2}(N_e-1)}{2 Z}{2\tau}.\nonumber
\end{eqnarray}
Here $s\in \{0,1\}$ labels the momentum sectors, so $K_x=sN_e$, and $z_{ij}=\frac{z_i-z_j}{L}$ and $Z=\sum_{j=1}^{N_e}\frac{z_i}{L}$ are the relative and center of mass coordinates respectively.
The modular parameter $\tau$ encodes the geometry of the torus and appears in the generalized Jacobi theta function
\[
\ellipticgeneralized a b z\tau = \sum_{k=-\infty}^\infty e^{i\pi\tau(k+a)^2}e^{i2\pi(k+a)(z+b)}.
\]
The function $\theta_1$
 which appears in the torus version of the Jastrow factor is defined as $\elliptic1z\tau=\ellipticgeneralized {\frac12}{\frac12}z\tau$.
Note that the index $s$ only appears in the factor which depends on the 
centre of mass coordinate $Z$, as is to be expected, since it sets the center of mass momentum. 

We can now write CF liquid wave functions for either choice of $s$ as
\[
\Psi_{K_1,K_2,s} = \det\left[t^{(i)}_{-k^{(j)}_{y},k^{(j)}_{x}}\right]\psi_{\frac{1}{2}}^{\left(s\right)}\left(z\right)\label{eq:Final_eqn_2}
\]
Here, the collection of $\mathbf{k}^{(j)}$ form some Fermi disc as usual and we ignore a scale factor which depends on the $\mathbf{k}^{(j)}$  present in Eq. ~\eqref{eq:CF_projected} (The functions $\Psi_{K_1,K_2,s}$ are not normalized anyway, with or without this factor.)
Global magnetic translations act on $\Psi_{K_1,K_2,s}$ as follows,
\begin{eqnarray}
T_{x}\Psi_{K_1,K_2,s} & = & e^{\i2\pi\frac{K_{1}+N_{e}s}{N_{s}}}\Psi_{K_1,K_2,s}\nonumber \\
T_{y}^{2}\Psi_{K_1,K_2,s} & = & e^{\i2\pi\frac{K_{2}}{N_{e}}}\Psi_{K_1,K_2,s}\nonumber \\
T_{y}\Psi_{K_1,K_2,s} & = & e^{\i2\pi\frac{K_{2}}{N_{s}}}\Psi_{K_1,K_2,s+1}.\label{eq:Traslations}
\end{eqnarray}
Here we have used $(K_1,K_2)$ to label the momentum of the composite fermions. The full momentum of the wave function $\Psi_{K_1,K_2,s}$ is then $(\tilde{K}_{1} ,\tilde{K}_{2})=(K_{1}+N_{e}s,K_{2})$. It appears that we can obtain different trial wave functions at a given value of $\mathbf{\tilde{K}}$ using either $K_1=\tilde{K}_{1}$ and $s=0$ or using $K_1=\tilde{K}_{1}-N_e$ and $s=1$. However, we now show that the wave functions with $s=1$ are actually equal to wave functions with $s=0$ where the momenta $k^{i}_{1}$ of all CFs are shifted by one unit. 

Consider the effect of a global change $k^{j}_{y}\to k^{j}_{y}-k_{0}$ for all $j$.
We have
\begin{eqnarray*}
 \det\left[t_{-k_{y}^{j}+k_{0},k_{x}^{j}}^{\left(i\right)}\right]
  &=&\det\left[e^{-\i2\pi\frac{k_{0}k_{x}^{j}}{2N_{s}}}t_{k_{0},0}^{\left(i\right)}t_{-k_{y}^{j},k_{x}^{j}}^{\left(i\right)}\right]\\
 & = & \prod_{i}t_{k_{0},0}^{\left(i\right)}e^{-\i2\pi\sum_{j}\frac{k_{0}k_{x}^{j}}{2N_{s}}}\det\left[t_{-k_{y}^{j},k_{x}^{j}}^{\left(i\right)}\right]\\
 & = & e^{-\i2\pi\frac{k_{0}K_{1}}{2N_{s}}}T_{x}^{k_{0}}\det\left[t_{-k_{y}^{j},k_{x}^{j}}^{\left(i\right)}\right]\\
\end{eqnarray*}
Similarly we have that if $k^{j}_{x}\to k^{j}_{x}+k_{0}$, then 
\[
\det\left[t_{-k_{y}^{j},k_{x}^{j}+k_{0}}^{\left(i\right)}\right]=
  e^{-\i2\pi\frac{k_{0}K_{2}}{2N_{s}}}T_{y}^{k_{0}}\det\left[t_{-k_{y}^{j},k_{x}^{j}}^{\left(i\right)}\right].
\]
Using $\eqref{eq:Traslations}$ we can write the effect on $\Psi_{K_1,K_2,s}$
as 
\begin{eqnarray*}
\Psi_{K_1,K_2+N_ek_0,s} & = & \Psi_{K_1,K_2,s}e^{\i2\pi\frac{k_{0}K_{1}}{2N_{s}}}e^{\i2\pi\frac{k_{0}N_{e}s}{N_{s}}}\\
\Psi_{K_1+N_ek_0,K_2,s} & = & \Psi_{K_1,K_2,s+k_0}e^{\i2\pi\frac{k_{0}K_{2}}{2N_{s}}}.
\end{eqnarray*}
Note here that $\Psi_{K_1+N_s,K_2,s}\propto\Psi_{K_1,K_2,s}$ and $\Psi_{K_1,K_2+N_e,s}\propto\Psi_{K_1,K_2,s}$.
The conclusion is thus that we are able to mod out $K_2\to K_2+ N_e$
and $K_1\to K_1+N_s$ but not $K_1\to K_1+N_e$.
Note however that  $\Psi_{K_1+N_ek_0,K_2,s} \propto \Psi_{K_1,K_2,s+k_0}$ which means that $\tilde K_1\to \tilde K_1+N_e$ can be implemented  by changing the $s$-sector.
This last result also tells us that any configuration with $K_1\to K_1+N_e$ is after projection equivalent to an unshifted configuration where $s\to s+1$.
This freedom can be used to minimize MC errors, by choosing the Laughlin state $s$ that minimizes $\sum_{j}\left|k_{j}\right|^{2}$ in \eqref{eq:CF_projected}.

\end{document}